\documentclass[manuscript]{emulateapj}
\usepackage{epsfig}
\usepackage{apjfonts}

\newcommand{\lsim}{\lower0.6ex\vbox{\hbox{$ \buildrel{\textstyle<}\over{\sim}\ $}}}
\newcommand{\gsim}{\lower0.6ex\vbox{\hbox{$ \buildrel{\textstyle>}\over{\sim}\ $}}}

\newcommand{\Vmax}{V_{\rm max}}
\newcommand{\Vin}{V_{\rm in}}
\newcommand{\Vnow}{V_{\rm now}}

\newcommand{\Msun}{M_{\odot}}
\newcommand{\hMsun}{{\rm h}^{-1} \Msun}
\newcommand{\hMpc}{{\rm h}^{-1} {\rm Mpc}}

\newcommand{\Rc}{R_c~}
\newcommand{\beq}{\begin{equation}}
\newcommand{\eeq}{\end{equation}}

\shorttitle{Galaxy Environments} 
\shortauthors{Berrier et~al.}

\begin{document}

\title{Counts-in-Cylinders in the Sloan Digital Sky Survey with Comparisons to N-body Simulations
}
\author{
Heather D. Berrier,
Elizabeth J. Barton
}
\affil{Center for Cosmology, Department of Physics and Astronomy,
The University of California at Irvine, Irvine, CA 92697, USA}
\author{Joel C. Berrier}
\affil{Department of Physics, University of Arkansas, 835 West 
Dickson Street, Fayetteville, AR 72701, USA \\
Arkansas Center for Space and Planetary Sciences, 202 Old Museum
 Building, University of Arkansas, Fayeteville, AR 72701, USA}
\author{James S. Bullock}
\affil{Center for Cosmology, Department of Physics and Astronomy,
The University of California at Irvine, Irvine, CA 92697, USA}
\author {Andrew R. Zentner}
\affil{Department of Physics and Astronomy, University of Pittsburgh, 
Pittsburgh, PA 15260, USA}
\author {Risa H. Wechsler}
\affil{Kavli Institute for Particle Astrophysics \& Cosmology, Department of Physics 
, and SLAC National Accelerator Laboratory, Stanford University,
Stanford, CA 94305, USA}

\begin{abstract}

Environmental statistics  provide a  necessary means of  comparing the
properties of galaxies  in different environments and a  vital test of
models  of  galaxy   formation  within  the  prevailing,  hierarchical
cosmological   model.   We   explore  counts-in-cylinders,   a  common
statistic defined as  the number of companions of  a particular galaxy
found within  a given projected radius and  redshift interval.  Galaxy
distributions  with the  same two-point  correlation functions  do not
necessarily have the same  companion count distributions.  We use this
statistic to examine the environments of galaxies in the Sloan Digital
Sky Survey, Data  Release 4.  We also make  preliminary comparisons to
four  models  for the  spatial  distributions  of  galaxies, based  on
$N$-body simulations, and  data from SDSS DR4 to  study the utility of
the  counts-in-cylinders statistic.   There  is a  very large  scatter
between the  number of  companions a  galaxy has and  the mass  of its
parent dark matter halo and  the halo occupation, limiting the utility
of this statistic for certain kinds of environmental studies.  We also
show that prevalent, empirical  models of galaxy clustering that match
observed  two-  and three-point  clustering  statistics  well fail  to
reproduce   some    aspects   of   the    observed   distribution   of
counts-in-cylinders on 1, 3 and  6-$\hMpc$ scales.  All models that we
explore  underpredict  the  fraction   of  galaxies  with  few  or  no
companions in 3 and 6-$\hMpc$ cylinders.  Roughly $7\%$ of galaxies in
the real universe  are significantly more isolated within  a 6 $\hMpc$
cylinder  than the  galaxies in  any of  the models  we  use.  Simple,
phenomenological models that map galaxies to dark matter halos fail to
reproduce    high-order   clustering    statistics    in   low-density
environments.
\end{abstract}


\section{Introduction} \label{sec:intro}

Measurements  of  galaxy  environments   provide  a  crucial  test  of
large-scale structure  and of the  physics of galaxy  formation.  Long
used    as    a    test    of   cosmological    models    \citep[e.g.,
][]{Blumenthal84,Bryan98,Bullock02,Berlind05,Berrier06,Blanton07},
environmental  statistics  become   more  powerful  probes  of  galaxy
formation models  as the cosmological  parameters of our  Universe are
measured with higher accuracy.

In  the modern  view of  galaxy formation,  galaxies form  within dark
matter halos. At any given epoch the relationship between galaxies and
their  dark matter  halos  can  be described  by  a ``halo  occupation
distribution'' (HOD),  which specifies the probability that  a halo of
mass $M$ hosts $N$ galaxies  with a given criteria.  This relationship
is still quite difficult to predict from first principles, and thus it
is useful  to use measurements of various  environmental statistics to
empirically constrain  this distribution  and to inform  more physical
models.  The  two-point correlation function  has long been  among the
most   powerful   tools   to   characterize   large-scale   structure,
\citep[e.g.,
][]{Peebles73,Kirshner79,Davis83,deLapparent88,Norberg01,Zehavi05,Padmanabhan07},
and has frequently  been used to constrain the HOD  for a given galaxy
population \citep{Scoccimarro01,Berlind02,Abazajian05,Lee06,Zheng07}.

A related way to probe  the distribution of galaxy environments is the
close-pair  fraction,  which has  been  used  widely in  observational
surveys to characterize the evolution of galaxy merger rates, enabling
tests of  the hierarchical merger  sequence predicted by  the standard
cosmological            model            \citep{Zepf89,Yee95,Patton97,
  Patton02,Lin04,DePropris05,DePropris07}.  \citet{Berrier06} examined
the evolution  of the close-pair fraction  of dark matter  halos in an
$N$-body simulation.   They find that the close-pair  fraction of dark
matter   halos  does  not   directly  measure   the  merger   rate  of
galaxies. However,  the predicted halo close-pair counts  do match the
observed  close-pair   fraction  of  galaxies,   assuming  that  every
sufficiently large  dark matter halo  contains a galaxy.   While these
results  are encouraging,  tests  in other  regimes  are necessary  to
assess both the underlying cosmological  model and the manner in which
galaxies are related to overdensities of dark matter.

The local number density of galaxies has a well-established connection
to  the   morphologies  and   colors  of  individual   galaxies;  this
relationship   is    known   as   the    morphology-density   relation
\citep{Oemler74, Dressler80, Postman84,Park07}.  There are several methods of
measuring density.   \citet{Dressler80} uses the  10 nearest neighbors
to calculate  the local surface  density.  More recently,  many groups
use  counts within spheres  \citep{Hogg03, Blanton03c,  Blanton05b} or
cylinders   \citep{Hogg04,Blanton06,Kauffmann04,Barton07}   of   fixed
radius.  As  with group finding  algorithms, this method  suffers from
both   incompleteness  and   contamination.   However,   it   is  more
straightforward  and well-defined  to implement  number counts  than a
group-finding algorithm.

In this  paper, we further explore the  utility of counts-in-cylinders
statistics as  a diagnostic of  galaxy formation models.   We consider
both    semi-analytic    galaxy    formation    models    (based    on
publicly-available catalogs from the Millennium simulation) as well as
methods based  on halo abundance  matching.  The latter type  of model
uses  high resolution dissipationless,  cold dark  matter simulations,
combined  with simple prescriptions  for the  galaxy--halo connection.
Such  models have  proved  remarkably successful  in matching  several
statistics  of  the   galaxy  distribution,  including  two-point  and
three-point                    correlation                   functions
\citep{Carlberg91,Colin97,Kravtsov04,Neyrinck04,Conroy06,Marin08}.

The  paper is  organized  as follows.   \S~\ref{sec:data} provides  an
overview of  our analysis of counts-in-cylinders in  the Sloan Digital
Sky  Survey  data.   In  \S~\ref{sec:Sims}  we  discuss  the  $N$-body
simulations and the phenomenological  models used in our analysis.  We
compare  the  observational and  simulation  results  as described  in
\S~\ref{sec:envstat}.     We    give    our   primary    results    in
\S~\ref{sec:results}.   First,   we  demonstrate  the  complementarity
between  the two-point  correlation  function and  counts-in-cylinders
statistics.  We  then use  the counts-in-cylinders statistic  to probe
galaxy  environments in  several models  for the  galaxy distribution.
Second, we show that galaxies in the SDSS sample are considerably more
isolated than  galaxies in  any of four  mock galaxy catalogs  that we
consider.

In  \S~\ref{sec:discussion},  we   discuss  our  results  and  explore
potential  systematic  issues  stemming  from  either  our  simulation
analysis or  our treatment  of the SDSS  sample.  We  draw conclusions
from our analysis in \S~\ref{sec:conc}.

\section{Observational Data: the Sloan Digital Sky Survey}
\label{sec:data}

We  use  the   Sloan  Digital  Sky  Survey  (SDSS),   Data  Release  4
\citep{Adelman06}.   Specifically, we  use the  Large  Scale Structure
subset of  the NYU-Value Added Galaxy Catalog  (NYU-VAGC), compiled by
\citet{Blanton05}.  The combined  spectroscopic sample in the NYU-VAGC
covers an area of 2627  square degrees, to an apparent magnitude limit
of $r =  17.77$.  Here we use the NYU-VAGC  to create a volume-limited
catalog of  27959 objects,  limited to M$_{\rm  r} < -19  + 5\log{h}$,
with redshift limits 0.0044 $\geq z \leq$ 0.0618.

Of  course,  the  SDSS   is  an  incomplete  redshift  survey.   Fiber
collisions cause  an estimated incompleteness of $\sim$  6\%, all from
pairs of galaxies  closer than 55$^{\prime\prime}$ \citep{Blanton03b}.
An  additional $\sim$  1\% of  galaxies are  missed because  of bright
foreground  stars.  As  described below,  we use  the  random catalogs
provided on the NYU-VAGC website,  which have the same geometry as the
survey,  to estimate  the  fraction of  companions  missed because  of
incompleteness, which we apply as  a correction to the cylinder counts
data.  We also use the random catalogs to determine where the cylinder
used  for our  companion counts  analysis falls  off the  edge  of the
survey.

\section{Simulations}
\label{sec:Sims}

We  compare  the SDSS  data  against  several  models for  the  galaxy
distribution,  based on  $N$-body simulations.   We use  two different
simulations, and  for each simulation we examine  two distinct methods
for matching galaxies to dark matter host halos and subhalos; thus, we
examine four distinct model galaxy catalogs, which we will refer to by
their  brief   names  (Z05   $\Vin$,  Z05  $\Vnow$,   Millennium,  and
MPAGalaxies)  as a  convenient  shorthand.  Each  of  these models  is
physically motivated,  though some are more strongly  favored.  We use
them to test  different physical models and to  account for the cosmic
variance we expect to find  in an SDSS-sized sample.  The simulations,
models,  and redshift  surveys used  in this  paper are  summarized in
Table~\ref{tab:t1}.   Please  note that  the  term  ``halos'' is  used
throughout  this work  to  mean  ``all halos''  (i.e.,  both host  and
subhalos).

\begin{table*}[t]
\caption{
Summary of Data and Simulations
}
\label{tab:t1}
\begin{center}
\begin{tiny}
\begin{tabular}{|l|l|l|l|l|}

\hline
Name& Type& Size ($(\hMpc)^3$)& Description& Reason Included\\
\hline

SDSS DR4& Redshift Survey& 3.034$\times 10^6$;& Observational data,& Data\\
 & & 4783 square degrees & incompleteness $\sim$6\%& \\
\hline
\hline
Z05 $\Vnow$& DM N-body+Semi-analytic Substructure& $(120)^3 $& Model using current subhalo $\Vmax$ as& Reduced resolution issues\\ 
 & & &  proxy for mass to assign luminosities& \\
\hline
Z05 $\Vin$& DM N-body+Semi-analytic Substructure& $(120)^3 $& Model using accreted subhalo $\Vmax$ as& Reduced resolution issues\\
 & & & proxy for mass to assign luminosities& \\
\hline
Millennium& DM N-body Simulation& $(500)^3 $& Use mass to assign luminosities& Large simulation used to calculate\\
 & & & & cosmic variance and check for\\
 & & & &systematic errors in $\Vnow$\\
\hline
MPAGalaxies& Millennium + Semi-analytic & $(500)^3 $& Use luminosities generated by model& Used to explore how more complicated\\ 
 & Galaxy Modeling& & & galaxy assignment affects distribution\\
\hline

\end{tabular}
\end{tiny}
\end{center}
\end{table*}


\subsection{N-body Simulation and Substructure}
\label{sec:Zentner}
The  primary  simulation  to which  we  compare  the  data is  a  high
resolution    $N$-body    simulation,    previously    described    in
\citet{Allgood06},  \citet{Wechsler06},  and  \citet{Berrier06}.   The
simulations  were performed  using an  Adaptive Refinement  Tree (ART)
$N$-body code \citep{Kravtsov97} with a cosmology of $\Omega_m$ = 0.3,
$h$ = 0.7,  and $\sigma_8$ = 0.9.  The  simulation consists of 512$^3$
particles in  a comoving  box of 120  $h^{-1}$ Mpc  on a side,  with a
particle  mass  of  $m_p  \simeq  1.07 \times  10^9$  $\hMsun$.   This
simulation is used  to identify host halos, i.e.,  halos whose centers
do not  lie within the virial  radius of larger halo.   The host halos
are complete down to virial masses of $M \simeq$ 10$^{10}$ $\hMsun$.

Substructure is  included using the  semi-analytic technique described
in  \citet{Zentner05}.  As described  in \citet{Berrier06},  we ignore
substructure  from the host  halos, and  replace it  with substructure
using the semi-analytic formalism.   The number of subhalos that merge
with   a  given   host   halo  is   determined   using  the   extended
Press-Schechter  formalism   \citep{Somerville99}.   Subsequently,  we
model  the  evolution of  merged  subhalos  including  both mass  loss
processes  and dynamical  friction.   We track  the  evolution of  the
subhalos until their maximum  circular velocities drop below $V_{\rm
  max} = 80$ km $s^{-1}$.   Adding this semi-analytic component to the
simulation removes inherent resolution limits, eliminating the problem
of  ``over-merging,''  which  is  of  particular  importance  for  the
enumeration of close pairs and companion counts we perform.

\begin{figure}[t!]
\epsscale{1.0}
\plotone{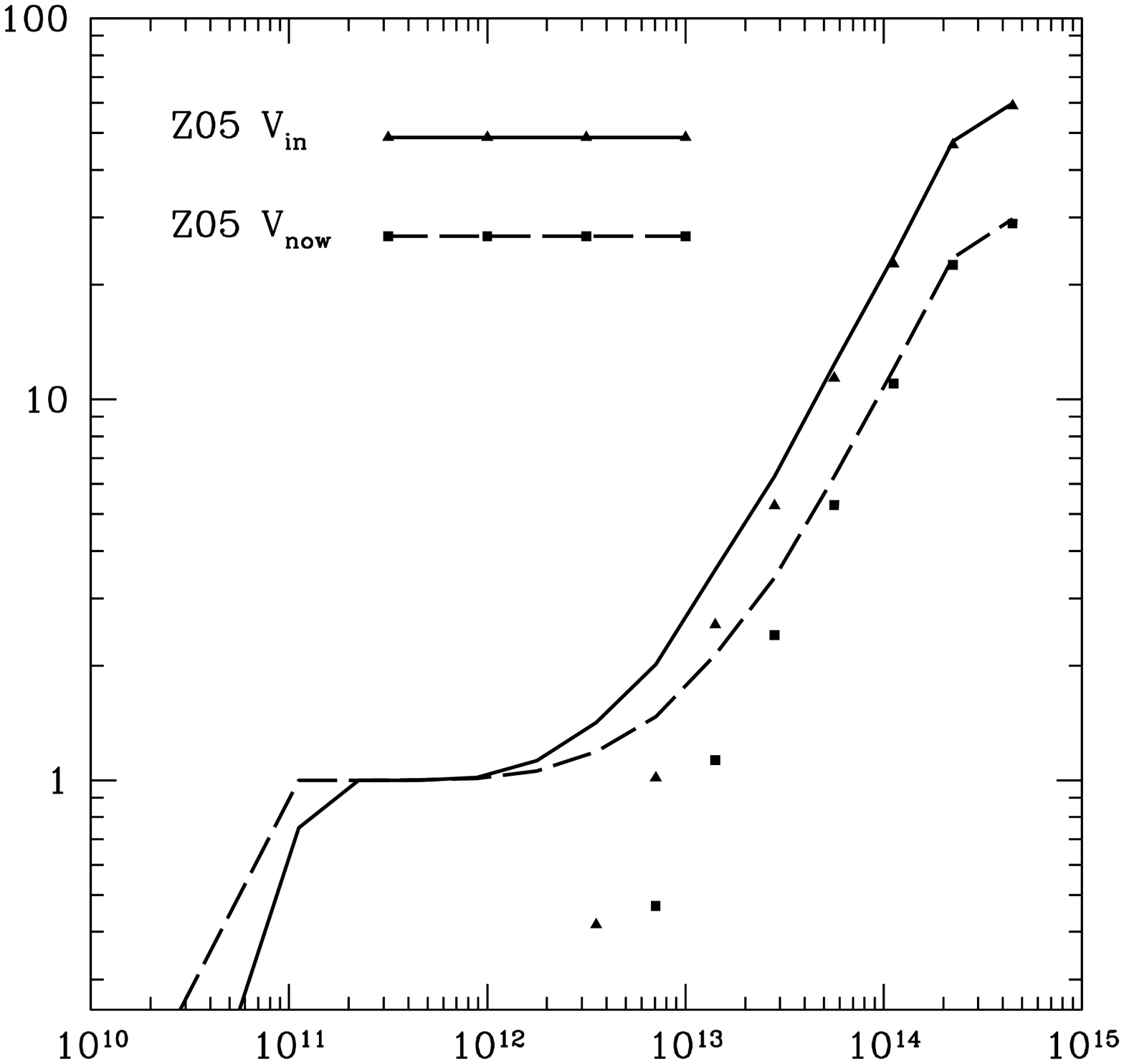}
\caption{ Halo occupation distribution  for the Z05 $\Vin$ and $\Vnow$
  models.   The  lines  are  the  combined HODs  for  host  halos  and
  subhalos, while the points are  the subhalos only.  The $\Vin$ model
  allows more subhalos on average to  survive in host halos of a given
  virial  mass (where the  virial overdensity  $\Delta_{vir}$ $\simeq$
  337). }
\label{fig:f1}
\end{figure}

We use the maximum circular velocity  of the halo, $\Vmax$, as a proxy
for  luminosity  in  what  follows, where  $\Vmax  \equiv  max[\sqrt{G
    M(<r)/r} ]$.   This measure is less ambiguous  than any particular
mass  definition.  Moreover,  subhalos  lose mass  at their  outskirts
rapidly upon  accretion, while halo  interiors (and thus  $\Vmax$) are
less severely affected  by this mass loss, so this  proxy is robust to
mild mass loss that likely would not affect the galaxy that resides at
the  subhalo center.   In this  work, we  use two  distinct  models to
compare  the simulation  (which  we  will refer  to  as ``Z05'')  with
observational data.  The first model uses the $\Vmax$ that the subhalo
has at the epoch of interest  (e.g., after it has been accreted by its
host halo,  evolved, and potentially lost significant  mass).  We will
refer   to  this   model   as  $\Vnow$.    Based   on  previous   work
\citep[e.g.,][]{Conroy06,Berrier06},   this  model  is   not  favored.
However, we note the results for the sake of completeness, and for the
sake of  comparison to the  Millennium simulation, which  also reports
the   $\Vmax$   at   the   epoch   of  interest   (as   described   in
\S~\ref{sec:Millennium}. The  second model, the  $\Vin$ model, assumes
that luminosity is related to the $\Vmax$ that the subhalo had just as
it was  being accreted and  prior to dynamical evolution.   The $\Vin$
model  has been shown  to reproduce  the galaxy  two-point correlation
function  at  many epochs  \citep{Conroy06},  as  well  as the  galaxy
close-pair fraction \citep{Berrier06} very well.  For subhalos $\Vnow$
is smaller than $\Vin$, because $\Vin$ characterizes subhalos prior to
the  removal of  mass by  the interaction  with the  host halo.   As a
result, the $\Vnow$  model has fewer subhalos above  any fixed maximum
velocity threshold.   The average number of  subhalos for a  host of a
given mass for  each of these models is  shown in Figure~\ref{fig:f1}.
These models are also described in more detail in \citet{Berrier06}.

\subsection{Identifying Galaxies with Halos}
\label{sec:sim2red}


To best  compare the simulations to  the data from SDSS,  we treat the
simulation  output as  if  it  were a  redshift  survey, by  assigning
luminosities to the dark  matter halos and restricting the information
used to that which would be obtainable from actual observations.

Since  the  N-body  substructure  simulations  contain  no  model  for
luminous  matter, we  use the  published $r$-band  luminosity function
from  the  SDSS  \citep{Blanton03},  and assign  luminosities  to  the
subhalos  to match  the observed  galaxy number  densities.  Following
\citet{Kravtsov04}, \citet{Conroy06} and \citet{Berrier06}, we assume a
one-to-one  relation  between dark  matter  halos (including  subhalos
within larger  host halos) and  galaxies.  Larger halos  correspond to
brighter  galaxies.  We establish  this relation  as follows:  for any
$r$-band magnitude,  we integrate  the luminosity function  to compute
the cumulative number density  of observed galaxies brighter than this
magnitude.   We  match  this  to  a halo  $\Vmax$  by  assigning  this
magnitude to the $\Vmax$ value for which the cumulative number density
of all halos is the same  as the cumulative number density of observed
galaxies.

For the Z05 simulation, we use both the $\Vin$ and $\Vnow$ models.  As
discussed  in \citet{Berrier06}, assigning  luminosities based  on the
$\Vnow$ model assumes that  baryons may be stripped significantly from
the galaxy  as it evolves  in the potential  of the larger  host, thus
gradually  reducing the galaxy  luminosity.  This  model underproduces
satellite galaxies in the  simulation.  Alternatively, using $\Vin$ to
calculate the number density assumes that the baryons are more tightly
bound than the dark matter, and that the luminous galaxy does not lose
significant stellar  mass after accretion  onto the larger  host.  The
true evolution is likely between these extremes.  The $\Vin$ model has
significant observational  support, but  we consider both  models here
for the purposes of comparison.

After the  luminosities are  assigned, we use  all halos  and subhalos
with M$_{\rm r} < -19 +  5\log{h}$ (which corresponds to $\Vmax$ = 137
km $s^{-1}$  for the $\Vin$  model and $\Vmax$  = 127 km  $s^{-1}$ for
$\Vnow$).  ``Moving'' the simulation out to a distance of 500 $\hMpc$,
to  cover the  appropriate  redshift space,  we  convert the  x, y,  z
coordinates to RA, Dec, and redshift.  We then compute the distance on
the sky in exactly the same way  as we do for the redshift survey.  We
employ  periodic boundary  conditions with  the simulations  to ensure
that the cylinder we are using never falls off the edge.

\subsection{Millennium Simulation}
\label{sec:Millennium}

The  Millennium  simulation  was  performed  with  the  GADGET-2  code
\citep{Springel05b} and follows the evolution of 2160$^3$ particles of
mass  $8.6  \times 10^8$  $\hMsun$  in a  box  500~$\hMpc$  on a  side
\citep{Springel05a}.  The  cosmology is comparable to that  of the Z05
simulation.   In particular,  the Millennium  simulation  cosmology is
spatially flat  with $\Omega_m$ = 0.25,  $h$ = 0.73,  and $\sigma_8$ =
0.9.   Halo  identification  was  performed  at  run-time  during  the
numerical simulation and the resultant  halo catalogs, which we use in
the present  study, are  publicly available.  The  stated completeness
limit   of   these   catalogs   is   $M   \ge   1.7   \times   10^{10}
h^{-1}$~M$_{\odot}$.  For  our purposes, the primary  advantage of the
Millennium simulation, and the  halo and galaxy catalogs produced from
it, is the large volume  compared with the Z05 models.  The Millennium
simulation volume  is $\sim 42$  times larger than  the volume-limited
SDSS sample we consider.

The Millennium Database does not  provide $\Vmax$ for all subhalos, so
we use  halo mass as  a proxy for  galaxy luminosity in this  case. We
assign  luminosities as  described in  \S~\ref{sec:sim2red},  where we
rank  simulated  halos  by  their  mass, observed  galaxies  by  their
$r$-band luminosities,  and map halo mass onto  luminosity by matching
the cumulative number densities at the mass and luminosity thresholds.
For field halos, this is  much like the $\Vnow$ model described above,
aside from  variations in $\Vmax$ at  fixed mass due  to variations in
the internal structures of  halos.  For subhalos, the relation between
$\Vmax$  and mass  may  be significantly  altered  and exhibit  larger
scatter due to the interactions  between the satellite objects and the
host  potentials.  We  assign luminosities  by the  abundance matching
method, rank  ordering both halos and  galaxies, so we  expect that in
broad terms this assignment should  be similar to the $\Vnow$ model in
so much as there is little difference between rank ordering by mass or
$\Vmax$.  We expect these assignments to be significantly different in
the central regions of host halos, where the influence of interactions
on subhalos is large.

For further comparison we  also use the MPAGalaxies database described
in \citet{DeLucia07}.  \citet{DeLucia07} use a semi-analytic technique
to model  the galaxies  associated with the  dark matter halos  of the
Millennium Simulation.  \citet{DeLucia07} model the stellar components
of  these  galaxies  using  the \citet{Bruzual03}  stellar  population
synthesis  model,  with the  \citet{Chabrier03}  IMF  and Padova  1994
evolutionary tracks.  Galaxy mergers  follow the mergers of their host
dark matter halos  until the halos fall below  the resolution limit of
the simulation.   At that  point, De Lucia  and Blaizot  calculate the
survival time  of the  galaxies using their  orbits and  the dynamical
friction time.  Mergers result in a ``collisional starburst'' modeled
by the prescription in \citet{Somerville01}.

\subsection{The Two-Point Correlation Function}
\label{sub:CF}

\begin{figure}[t!]
\epsscale{1.0}
\plotone{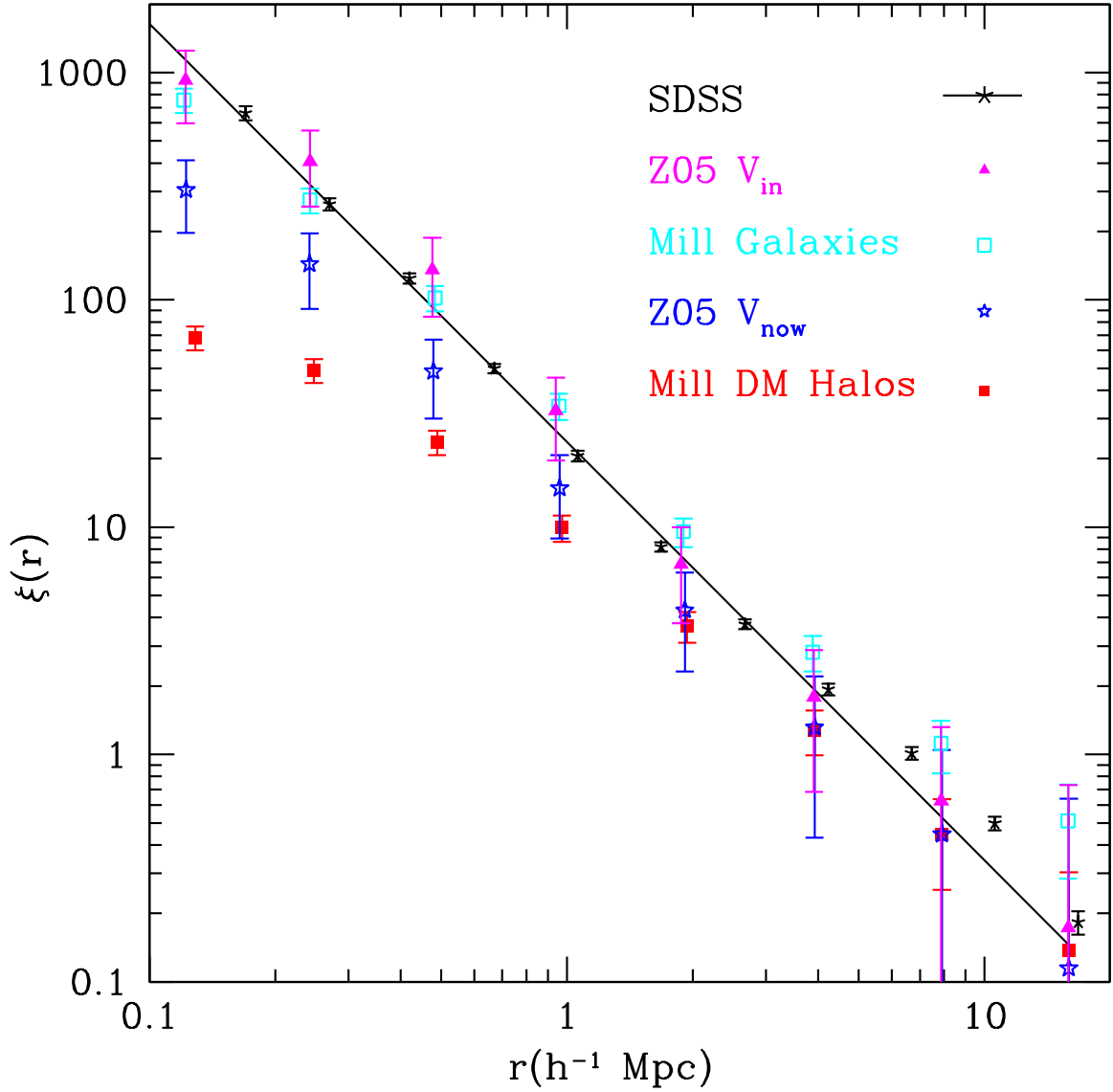}  
\caption{ Two-point  correlation functions  for models and  data.  The
  straight line  is the power  law fit to SDSS  from \citet{Zehavi05},
  $\xi (r) = (r/r_0)^{-\gamma}$, with  $r_0$ = 5.59 $\pm$ 0.11 $\hMpc$
  and $\gamma$ = 1.84 $\pm$  0.01, while the black stars represent the
  actual  data  (I.   Zehavi,  private  communication).   The  magenta
  triangles are  the correlation function  from the $\Vin$  model, the
  blue  stars  are  from  $\Vnow$,  the  solid  red  squares  are  the
  Millennium simulation  dark matter halos, and the  open cyan squares
  are  the  Millennium  simulation   galaxies.   The  error  bars  are
  calculated by jackknifing over octants of the simulation for the Z05
  models and over similarly sized subvolumes of the millenium sample. }
\label{fig:f2}
\end{figure}

As   stated   in   \S~\ref{sec:intro},   for   several,   simple   and
well-motivated assignments of galaxy  luminosity to dark matter halos,
predicted  two-point correlation  functions  match observations  well,
particularly on  large scales ($  \gtrsim $~Mpc).  Figure~\ref{fig:f2}
shows  the two-point  functions of  the simulated  galaxy  catalogs we
consider  compared to  the SDSS  measurement of  the  galaxy two-point
function  by  \citet{Zehavi05}.   In  broad terms,  the  $\Vin$  model
corresponds most  closely to  the SDSS data  over the entire  range of
scales,  as expected  from the  previous results  of \citet{Conroy06}.
The  $\Vnow$ and  Millennium  DM models  deviate  at relatively  small
separations.

The discrepancy seen at  small separations between the Millennium dark
matter  halos and  the  other  mock galaxy  samples  may have  several
causes.  Most likely, this reflects the fact that the relation between
halo  $\Vmax$ and  remaining bound  mass is  significantly  altered in
dense environments  relative to the  field.  It is also  possible that
the  simulation  could  suffer  from numerical  dissolution  of  small
subhalos in dense environments.   The MPAGalaxies sample is consistent
with the SDSS result above r $\cong$ .3 $\hMpc$.

\section{The Counts-in-Cylinders Environment Statistic}
\label{sec:envstat}

\subsection{General Description}
\label{sec:descrip}

The counts-in-cylinders  statistic that we  use is similar to  the one
used by  \citet{Kauffmann04} and \citet{Blanton06}.  We  look at every
halo or  galaxy in the  catalog and count  how many companions  it has
within  a  cylinder  with  a  radius defined  by  a  given  transverse
separation  and a depth  given by  a specified  line-of-sight velocity
difference.   We choose  a  cylinder  depth that  is  large enough  to
include all physically-associated galaxies, except in the most massive
clusters. Specifically, we  calculate the counts-in-cylinders for each
galaxy using four different radii  --- $\Rc$ = 0.5 $\hMpc$, 1 $\hMpc$,
3 $\hMpc$,  and 6 $\hMpc$  --- searching for companions  with velocity
differences of $| \Delta$V | $\le$ 1000 km s$^{-1}$.

\subsubsection{Correcting for Incompleteness in the SDSS Data} 
\label{sec:cv}

Using the same technique, we measure the counts-in-cylinders statistic
for galaxies  in our volume-limited  SDSS NYU-VAGC sample.   We search
each galaxy for  all companions within $\Rc$ and  $\Delta$V. The tools
of  the  NYU-VAGC  allow  us  to estimate  the  completeness  in  this
cylinder.   Specifically, we  search four  of the  random  catalogs of
galaxies  evenly distributed  throughout the  survey footprint  in the
apparent magnitude range and separation on the sky.  We weight each of
the  random  galaxies by  the  completeness  of  the sector  from  the
``lss$\_$geometry.dr4.fits FGOTMAIN'' parameter and by estimating what
is  missing  from  the  limiting  magnitude  of  the  sector  and  the
luminosity function  of the survey  \citep{Blanton05}.  After applying
this  weight,  we add  the  number  of  weighted random  galaxies  and
normalize by  the area searched  on the sky.   We then use  the random
counts  as  a  measure  of  the  local  completeness  of  the  survey,
normalizing  it  to  the mode  for  the  survey  and dividing  by  the
completeness to  arrive at an estimate  of the corrected  counts for a
particular    galaxy.    When    constructing   histograms    of   the
counts-in-cylinders statistics,  we weight each galaxy  by the inverse
of its corresponding completeness to account for missing {\it central}
(searched) galaxies in the survey.

Our correction  for incompleteness  undercounts the effects  of missed
close pairs, which are significantly more likely than ``random'' to be
coincident in redshift  space.  We test the effects  of these pairs by
constructing an  artificial simulation in  which we eliminate  80\% of
the close  pairs that  would appear closer  than 55 arcseconds  on the
sky, as  assigned by  a redshift distribution  that matches  the data.
The effect is not systematic  except at low companion counts, but even
there the  magnitude of  the effect averaged  over <= 5  companions is
$\lesssim 5\%, 10\%$,  and $12\%$ for the 1, 3,  and 6 $\hMpc$ scales,
respectively.

\subsection{Sample Variance} 
\label{sec:cv}

\begin{figure}[t!]
\epsscale{1.0} 
\plotone{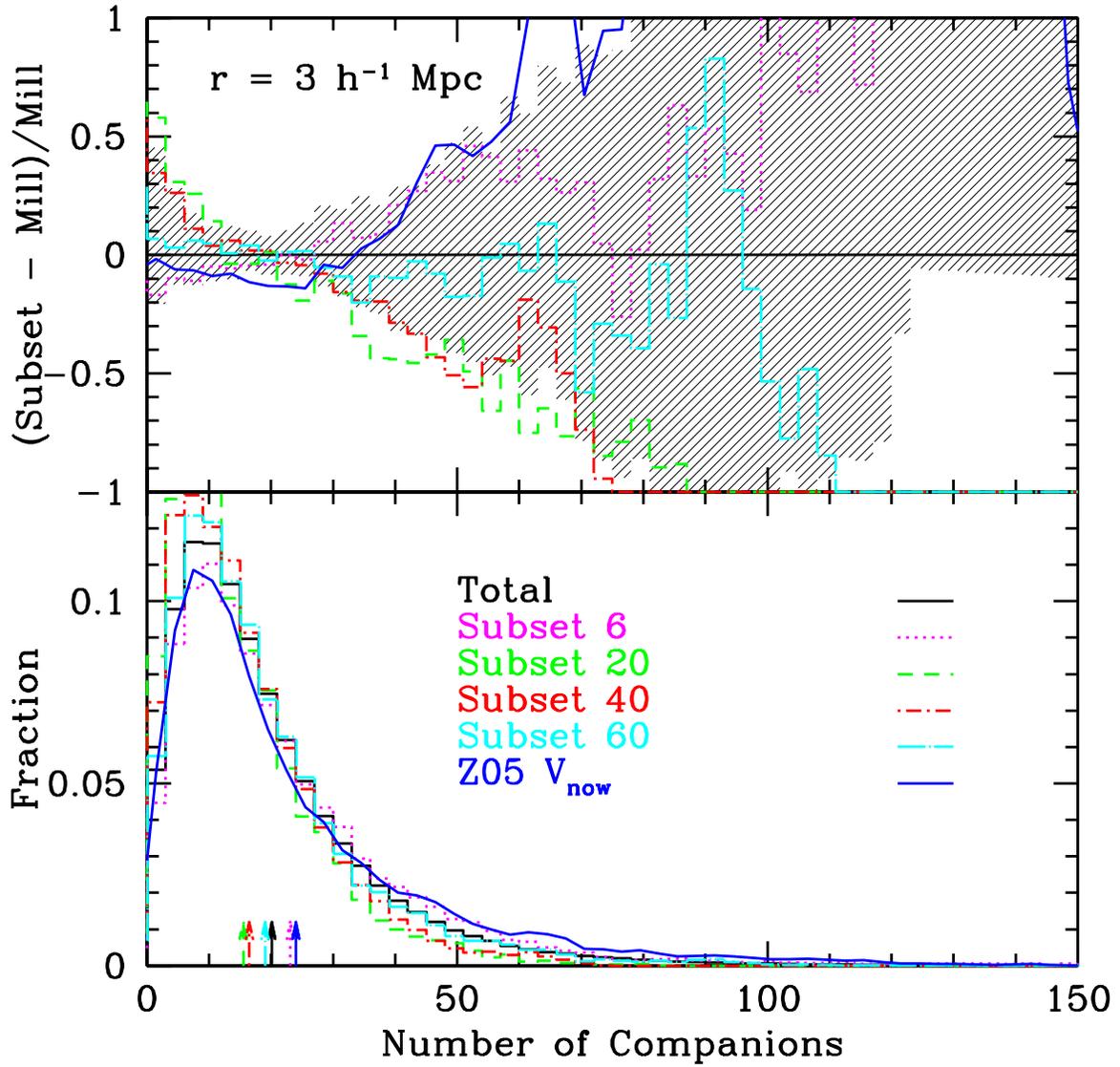}
\caption{  Comparison of  counts-in-cylinders  between the  Millennium
  simulation and  the Z05 model (smooth  blue line) for  the 3 $\hMpc$
  cylinder.  {\bf Bottom  panel:} Distribution of counts in  $\Rc$ = 3
  $\hMpc$ cylinders.  The arrows denote the mean of each distribution.
  {\bf Top panel:} Fractional deviation from the mean cylinder counts.
  The shaded area denotes the dispersion among the $64$ sub-volumes of
  the  Millennium   simulation  and  provides  a   guideline  for  the
  statistical limitations of the comparison.  }
\label{fig:f3}
\end{figure}

By  subdividing  the Millennium  simulation  volume,  we estimate  the
expected  sample  variance among  volumes  of  the  size of  the  SDSS
volume-limited sample  or the  Z05 simulation box  to ensure  that any
discrepancy  we  see is  larger  than  can  be attributed  to  natural
variations in large scale structure.  We calculate the cylinder counts
in 64 sub-volumes from the Millennium simulation.  Each sub-volume was
cubic with a side length of 125 $\hMpc$ and had a volume comparable to
our  Z05 catalogs.   The  histograms in  Figure~\ref{fig:f3} show  the
cylinder counts  for four of these  volumes for the $\Rc$  = 3 $\hMpc$
cylinder  (bottom),  and  the  variation  from  the  total  Millennium
distribution (top).   The errors  are 68\% from  the scatter  within a
bin.   We  do a  similar  calculation  for  SDSS-sized volumes  within
Millennium.  Those  results determine the errors on  the figures which
include SDSS data.

The  smooth blue  line overlaid  in Figure~\ref{fig:f3}  shows the
cylinder counts  result from  the Z05 $\Vnow$  model.  We  use $\Vnow$
because, as mentioned in \S~\ref{sec:Millennium}, this is the model to
which  the  Millennium  simulation  results  should  be  most  closely
related.  The $\Vnow$ result falls  for the most part within the error
bars  of the  Millennium distribution,  with a  $\chi^{2}$/(degrees of
freedom [dof])  value of 0.609.   The slight disagreement at  the peak
may be the  result of a non-trivial relation  between mass and $\Vnow$
in  dense  regions, an  inherent  shortcoming  in  the Z05  model,  or
numerical overmerging and halo incompleteness in dense environments in
the Millennium simulation.  The first  of these options is expected on
physical grounds and seems most likely.  The results for the $\Rc$ = 1
and 6 $\hMpc$ cylinders show similar trends.

\subsection{Companions as a Function of Host Halo Mass and Halo Occupation} 
\label{sec:halo mass}

\begin{figure}[t!]
\epsscale{1.0}
\plotone{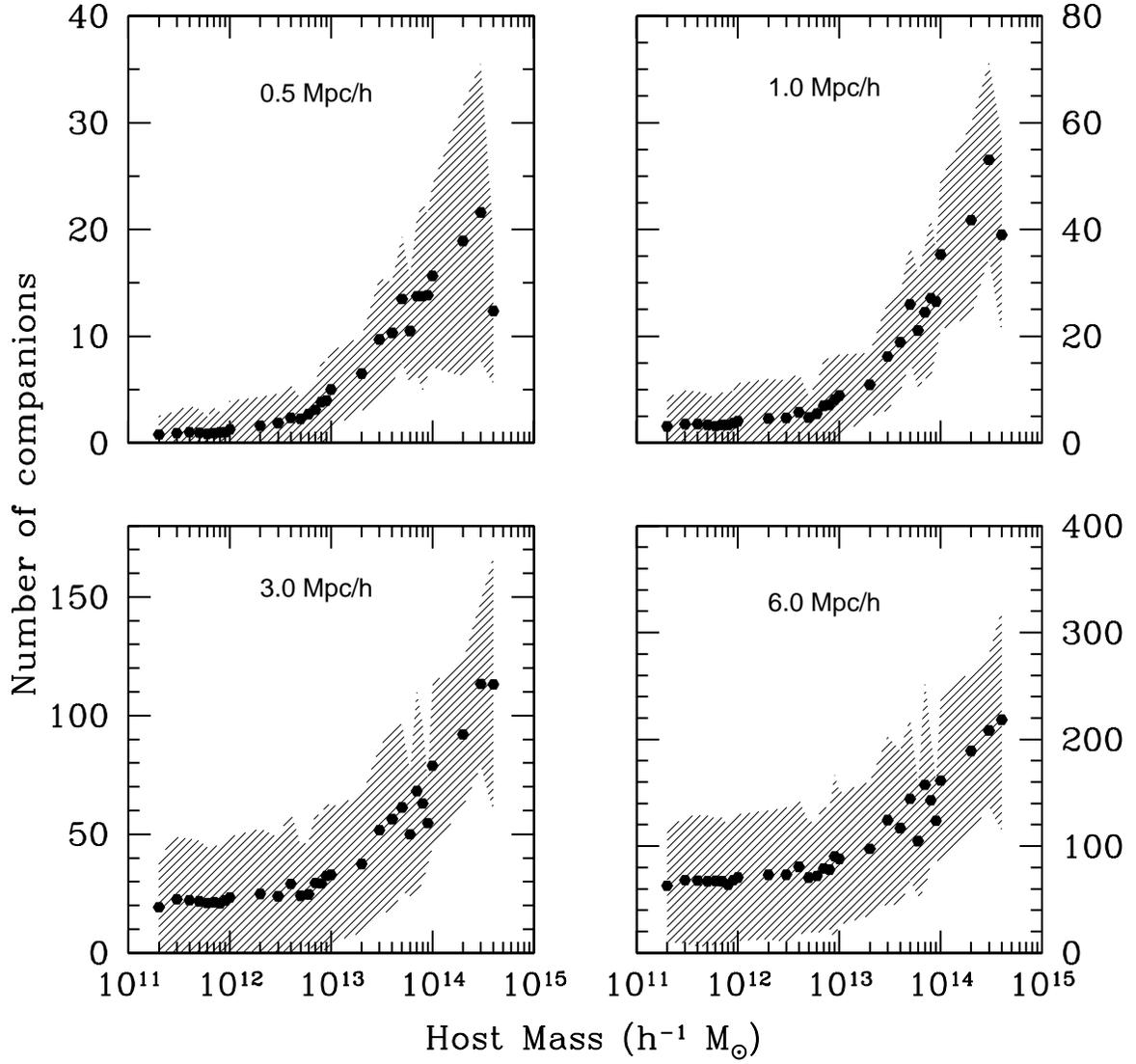}
\caption{ {\bf Clockwise from top left:} Average number of companions in
the Z05 $\Vin$ sample within $\Rc$ = .5, 1, 6, and 3 $\hMpc$ cylinders
for galaxies within a given  host halo mass bin.  Shaded regions indicate
the 1-$\sigma$ scatter in the value for each bin. Note that the y-axes
have different scales.}
\label{fig:f4}
\end{figure}

To  explore  the  utility  of the  counts-in-cylinders  statistic,  we
consider it  as a potential  proxy for the  masses of the  dark matter
host  halos  in  which  the  galaxies  reside.   Various  studies  use
counts-in-cylinders  to  test the  halo  model  and other  environment
predictors (e.g.,  Blanton et al. 2006).   We use the  $\Vin$ model to
calculate the  average number of  companions per galaxy within  a host
halo mass bin.  Figure~\ref{fig:f4} shows  the result for $\Rc$ = 0.5,
1,  3  and 6  $\hMpc$  cylinders.  The  error  bars  are one  standard
deviation    from   the    distribution   within    the    bin.    The
counts-in-cylinders statistic tracks mass and halo occupation, but the
relationships are  extremely noisy.  The $\Rc$ =  0.5 $\hMpc$ cylinder
can  potentially  distinguish  between  $\sim$  10$^{12}$  and  $\sim$
10$^{14}$ $\Msun$ halos.  The $\Rc$ = 1 $\hMpc$ cylinder distinguishes
10$^{12}$ $\Msun$  from 10$^{13}$ $\Msun$ and  10$^{14}$ $\Msun$.  The
$\Rc$ =  3 and 6 $\hMpc$  cylinders only distinguish  between halos at
the extremes  of the mass  distribution.  None of the  cylinders works
well for  masses below 10$^{12.5}$  $\Msun$, although this  scale will
vary with the  magnitude limit of the sample.   For masses this small,
the cylinders  frequently include multiple  small, physically-distinct
groups and clusters, rather than  solely the subhalos within one large
host.   For  large  companion   numbers,  the  smaller  cylinders  are
ineffective  because they  often do  not encompass  the  entire group,
which may have a virial radius  of $\sim 1 \hMpc$.  Ideally, one would
tune the cylinder radius and depth to the halo sample of interest.

The trends in the  1 and 6 $\hMpc$ scales may relate  to the result of
\citet{Blanton06}.   Blanton et  al.  show  that galaxy  color depends
much  more  strongly  on  the   1  $\hMpc$  density,  as  measured  by
counts-in-cylinders, than it does on the 6 $\hMpc$ surrounding density.

\begin{figure}[t!]
\epsscale{1.0}
\plotone{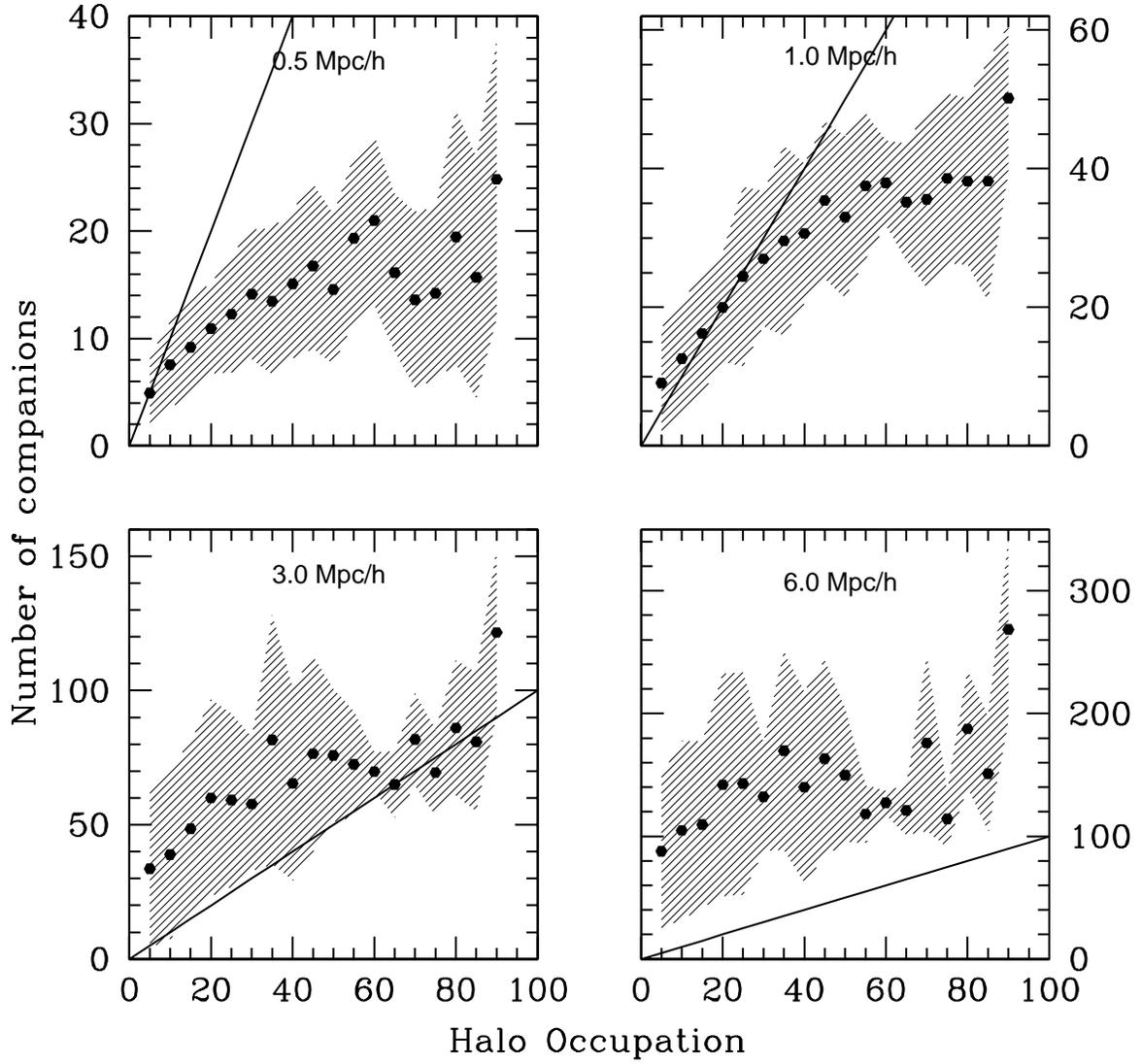}
\caption{ {\bf Clockwise from top left:}  Average number of companions in
the Z05 $\Vin$ sample within $\Rc$ = .5, 1, 6, and 3 $\hMpc$ cylinders
for galaxies  of  a  given  halo  occupation.   Solid  lines  correspond to   (halo
occupation) =  (number of  companions) for comparison.  Shaded regions
indicate the 1-$\sigma$ scatter in the value for each bin. }
\label{fig:f5}
\end{figure}

In  Figure~\ref{fig:f5},  we  explore  the  relationship  between  the
average number of companions in a given cylinder and the actual number
of subhalos residing within the host (the halo occupation).  The solid
lines  in the  four  panels represent  the  case where  the number  of
companions  is equal  to the  halo occupation.   We see  that  using a
cylinder of  $\Rc$= 1.0$\hMpc$  gives us a  very good estimate  of the
halo  occupation for  host halos  with fewer  than  $\sim$40 subhalos,
while a  cylinder of $\Rc$= 3.0  $\hMpc$ is reasonable  for host halos
with $\sim$65-85 subhalos.

\subsection{The Complementarity of Cylinder Counts and $N$-point Statistics}
\label{sec:rel}

The cylinder  count statistic is related to  the correlation function.
However, there  are important differences.   The two-point correlation
function  describes the  probability  of finding  companions within  a
spherical shell of  a given radius from a  galaxy, and the three-point
correlation  function  does  the  same  for three  galaxies  at  fixed
separations  from  each other,  and  so  on  for the  other  $N$-point
correlators.  The  cylinder-counts distribution gives  the probability
of finding a  certain number of companions within a  cylinder of a set
radius  from  a given  galaxy.   While  the  {\it average}  number  of
companions at the scale $\Rc$ is  set by the integral of the two-point
correlation  function  over  the  volume  of the  cylinder,  the  {\em
distribution}  of  companion counts  is  {\it  not}  specified by  the
two-point function.  For example, the variance of the companion counts
depends  upon  the three-point  function,  the  skewness of  companion
counts depends upon the four-point function, and so on.  In principle,
the {\em distribution} of companion  counts is sensitive to all of the
$N$-point  correlators   and  can  be  an  efficient   way  to  access
information not available through  two- and three-point statistics (or
through  the mean  of the  companion counts)  without  undertaking the
challenging task of computing higher-point correlation functions.

To illustrate the utility of cylinder-count statistics as a complement
to the correlation function, we identify two distributions of galaxies
that  effectively   yield  the  same  two-point   function,  but  have
systematically  different  companion   number  distributions.   As  an
example, we create a simple,  toy galaxy distribution by modifying the
Z05 $\Vin$ catalog.  We rearrange  the substructure so that host halos
that contain at  least one subhalo with 27-37  companions are stripped
of  all  of their  subhalo  companions  within  a 3  $\hMpc$  cylinder
(leaving  the   total  mass  unchanged).   These   subhalos  are  then
reassigned to  a location  within 1  $\hMpc$ of a  host halo  with >37
companions.    While   this   exercise   has  no   explicit   physical
justification,   it   does  create   a   test   catalog  where   halos
preferentially   avoid   environments   with  roughly   $\sim   30-40$
companions.

\begin{figure*}[t!]
\centering
\epsscale{1.0}
\plottwo{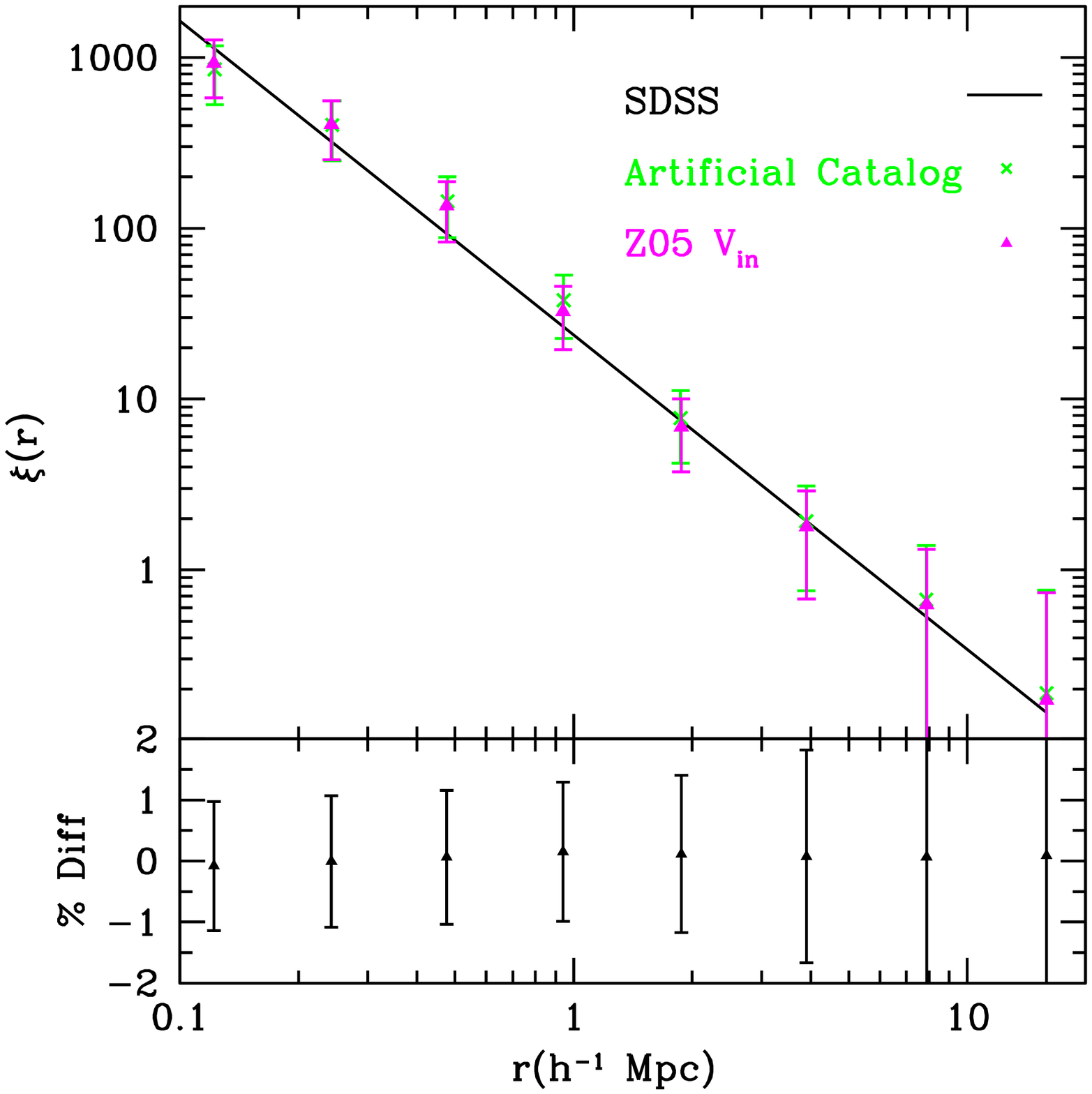} {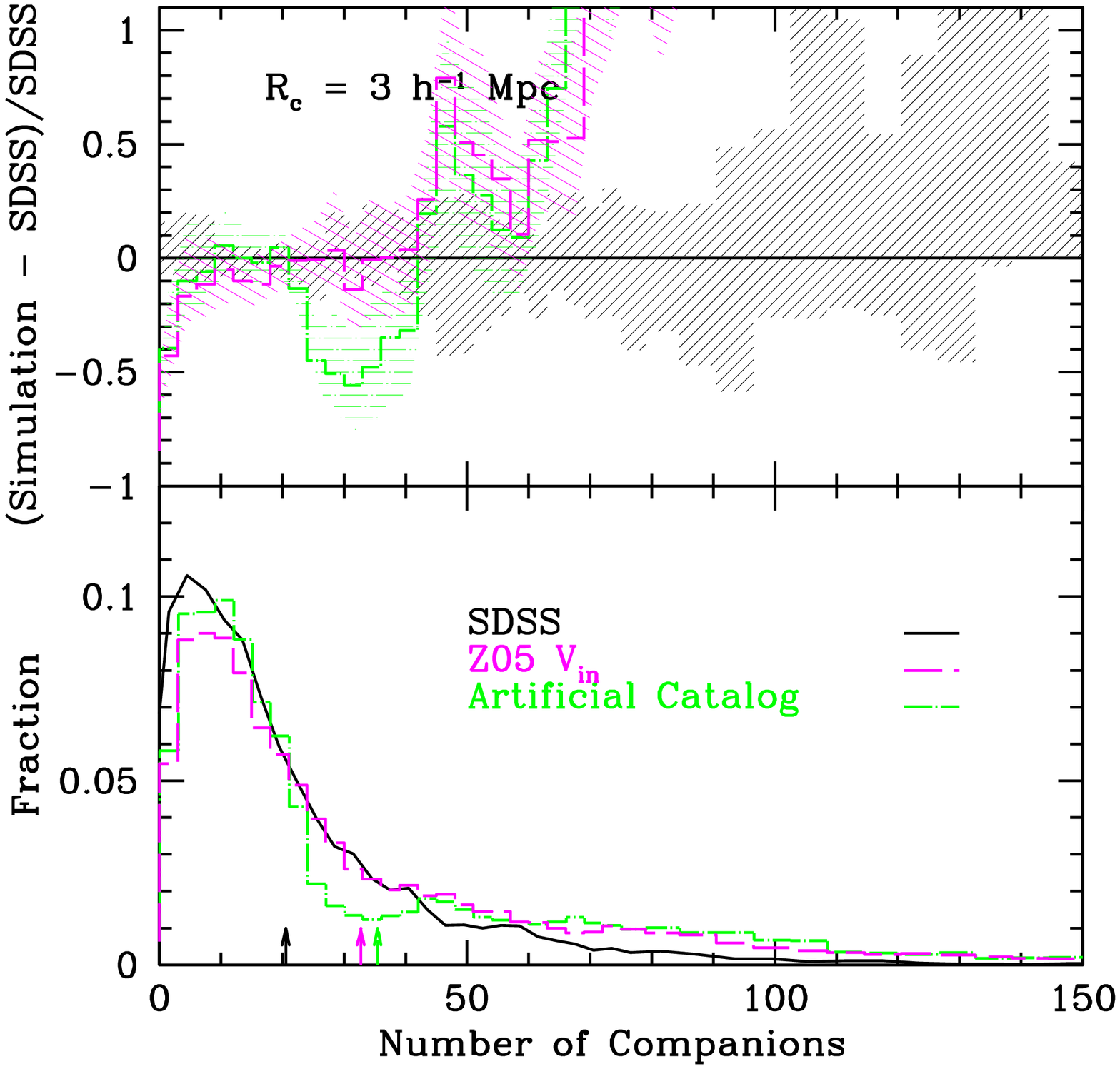}
\caption{{\bf  Top  left:}  Two-point  correlation  function  for  the
artificial catalog  (green ``x''s) and  the Z05 $\Vin$  model (magenta
triangles)  with the  best fit  line from  SDSS. Error  bars represent
jackknife  plus  Poisson  errors.   {\bf  Bottom  left:}  the  percent
difference between  the artificial catalog and the  $\Vin$ model. {\bf
Right:} Histogram  of the fraction of  galaxies or halos  with a given
number of companions within an $\Rc$ = 3 $\hMpc$ cylinder for the test
catalog  (green  long-dash-short-dash   line),  the  Z05  $\Vin$  data
(magenta dashed line) and SDSS (black solid line).  }
\label{fig:f6}
\end{figure*}

The  left-hand  panel   of  Figure~\ref{fig:f6}  shows  the  two-point
correlation function  of our  toy catalog compared  with the  best fit
line  from  SDSS  referenced  earlier and  the  two-point  correlation
function  from the  Z05 $\Vin$  model.  The  test  catalog correlation
function falls  well within the  error bars of the  $\Vin$ correlation
function.

In  contrast,  the  right-hand  panel of  Figure~\ref{fig:f6}  is  the
histogram  of the number  of companions  within an  $\Rc$ =  3 $\hMpc$
cylinder for the test catalog,  the $\Vin$ model and SDSS.  The figure
shows that the  effects of the substructure reassignment.   There is a
noticeable  dip  in the  test  catalog  data  right around  the  mean,
although  the mean itself  does not  change significantly.   Thus, the
cylinder  counts  statistic  yields  different information  about  the
distribution of substructure on the scale of the cylinder.

Notably,   this  tool   gives  more   direct  information   about  the
distribution  of  substructure  rather  than the  physical  separation
between objects.  It complements the 2-point correlation function as a
tool for  determining the accuracy  of clustering in  simulations when
compared to redshift surveys.


\section{Results} 
\label{sec:results}

\subsection{Comparing the Models to the SDSS} 
\label{sec:counts}

We now use the cylinder counts distribution described above to compare
the  predictions  from  the  simulation  catalogs  to  the  SDSS.   We
calculate  1-$\sigma$ uncertainties from  the cosmic  variance between
SDSS-sized volumes  within the Millennium Simulation to  set the error
bars for the SDSS data,  and use the 1-$\sigma$ uncertainties from the
cosmic variance  between Z05-sized volumes  as the error bars  for the
Z05  $\Vin$ and $\Vnow$  distributions.  We  summarize our  results in
tables~\ref{tab:t2} and ~\ref{tab:t3}.

\begin{figure*}[t!]
\centering
\epsscale{1.0}
\plottwo{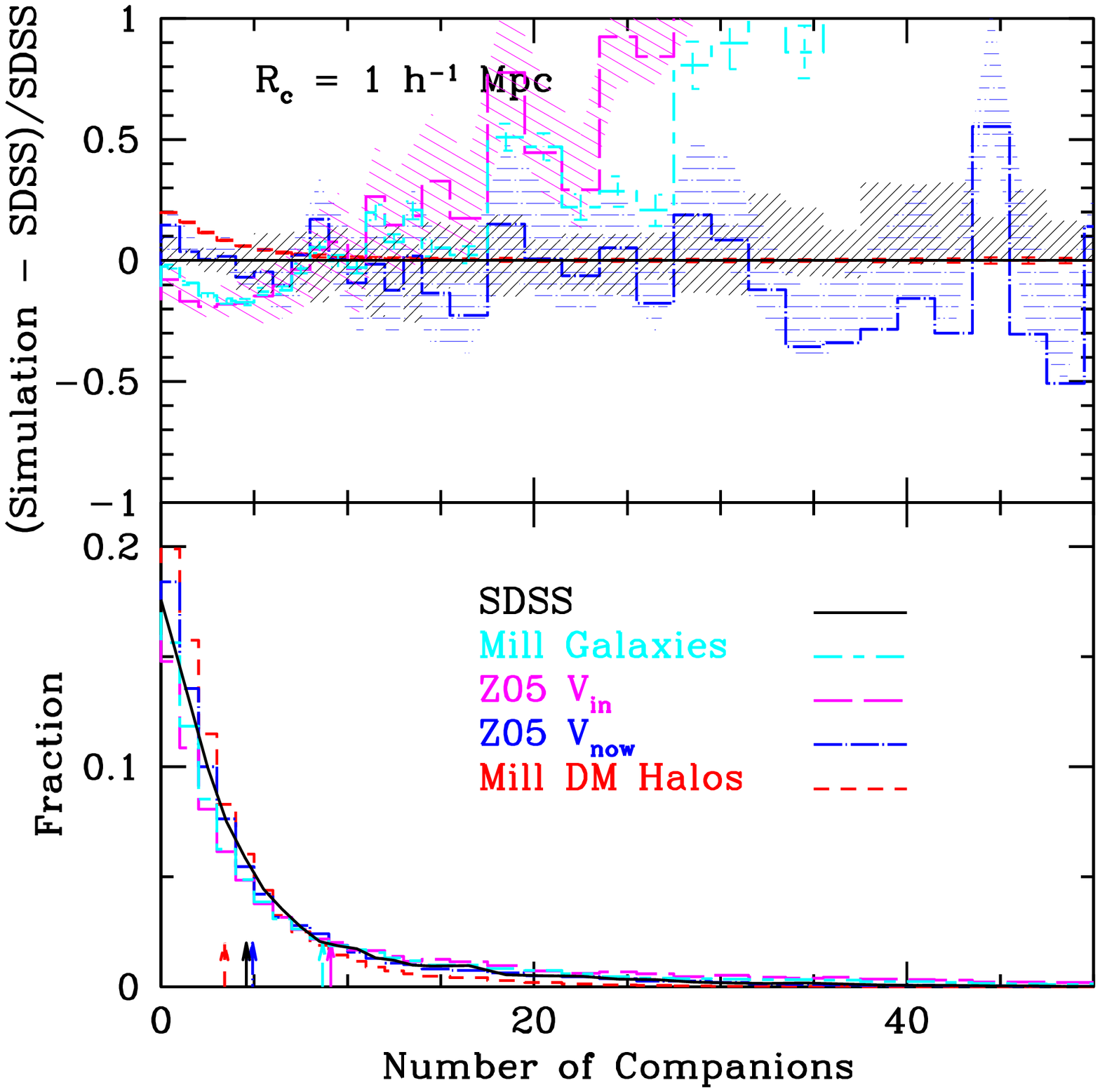} {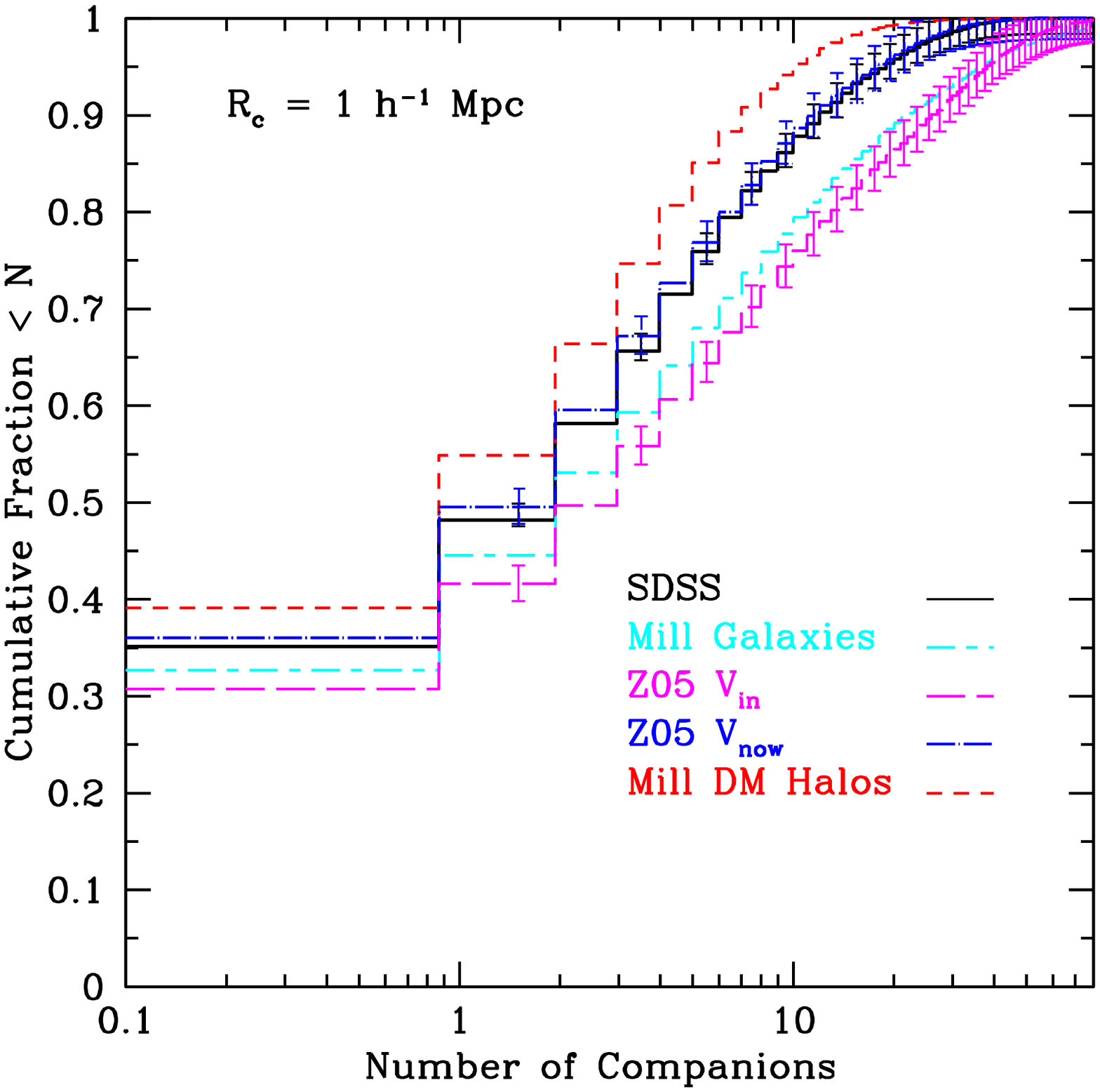}
\caption{ {\bf Bottom left:} Histogram  of the fraction of galaxies or
  halos with a given number of companions within the $R_c$ = 1 $\hMpc$
  cylinder for  Z05 $\Vin$  (magenta long-dashed line),  $\Vnow$ (blue
  dot-dashed  line), Millennium  (red short-dashed  line), MPAGalaxies
  (cyan short-dashed-long  dashed line), and SDSS  (smooth solid black
  line).  The  arrows show the  average number of companions  for each
  distribution.   {\bf  Top  left:}  Cylinders counts  with  the  SDSS
  distribution subtracted.  The shaded  area is 1-$\sigma$ from cosmic
  variance between  SDSS-sized volumes,  while the error  bars include
  the  cosmic   variance  between  Z05-sized   volumes.  {\bf  Right:}
  Cumulative  fraction of  of galaxies  or halos  with fewer  than the
  given number of companions within the $R_c$ = 1 $\hMpc$ cylinders.}
\label{fig:f7}
\end{figure*}

Figure~\ref{fig:f7} shows the counts-in-cylinders distribution
for the  $\Rc$ = 1 $\hMpc$  cylinder on the left,  and the cumulative
fraction  of galaxies or  halos with  fewer than  the given  number of
companions  on  the right.   The  arrows  denote  the mean  number  of
companions for each model.

\begin{figure*}[t!]
\centering
\epsscale{1.0}
\plottwo{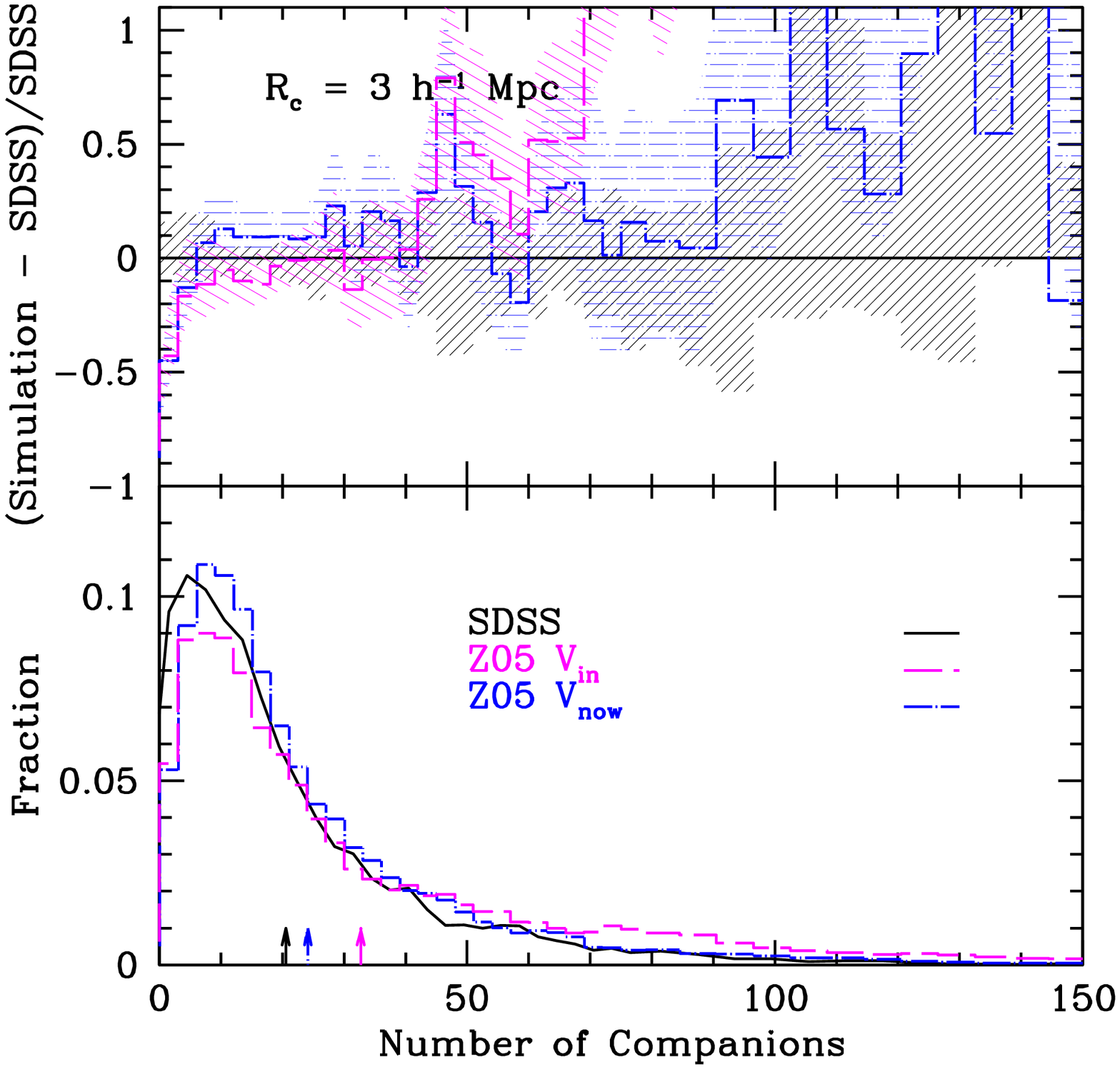} {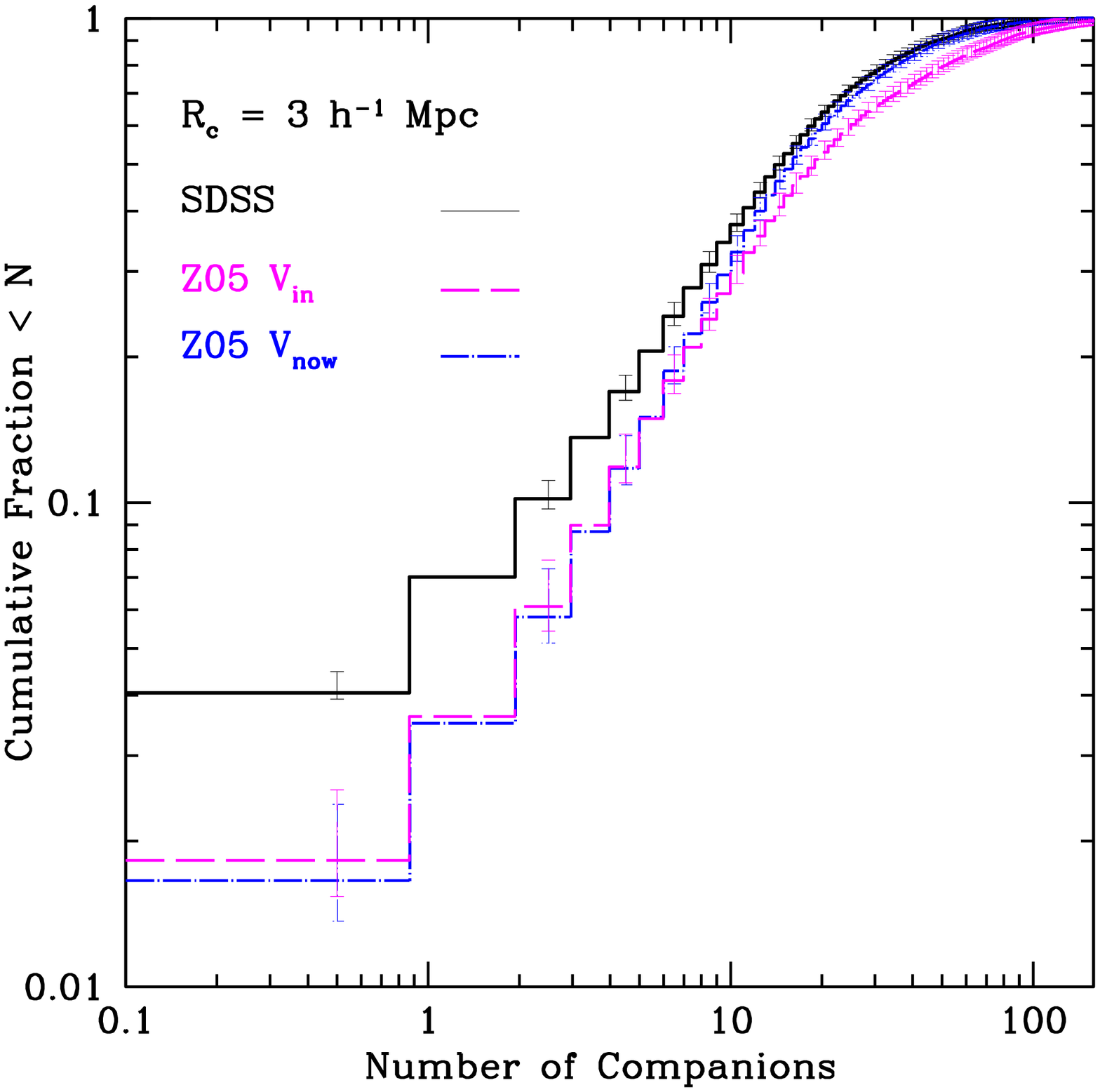} 
\caption{ {\bf  Left:}  Cylinder   counts  for  Z05  $\Vin$  (magenta
long-dashed line),  $\Vnow$ (blue  dot-dashed line), and  SDSS (smooth
solid black  line) within $\Rc$  = 3 $\hMpc$ cylinders.   {\bf Right:}
Cumulative fraction of of galaxies  or halos with fewer than the given
number of companions within $\Rc$ = 3 $\hMpc$ cylinders. }
\label{fig:f8}
\end{figure*}

\begin{figure*}[t!]
\centering
\epsscale{1.0}
\plottwo{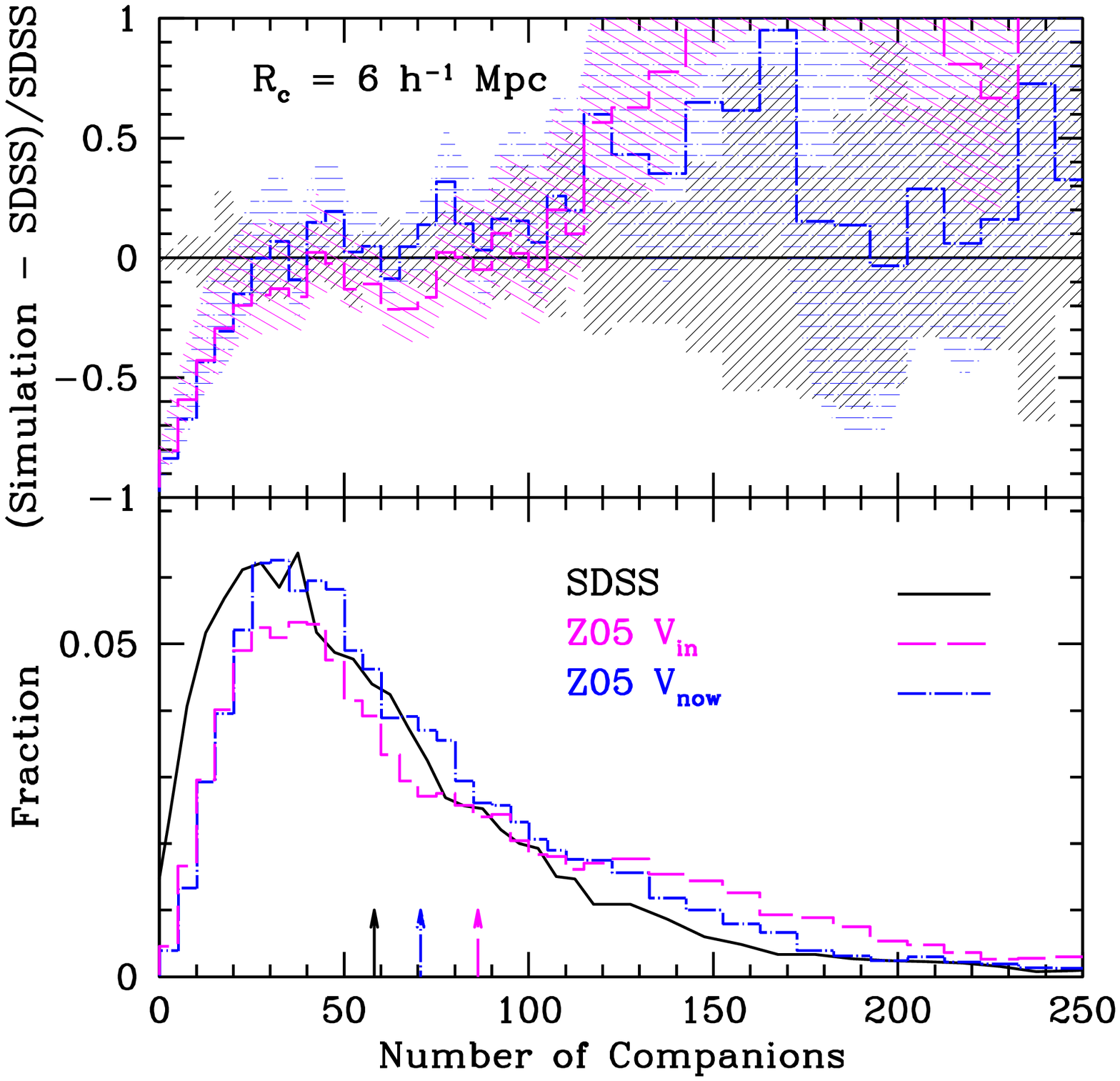} {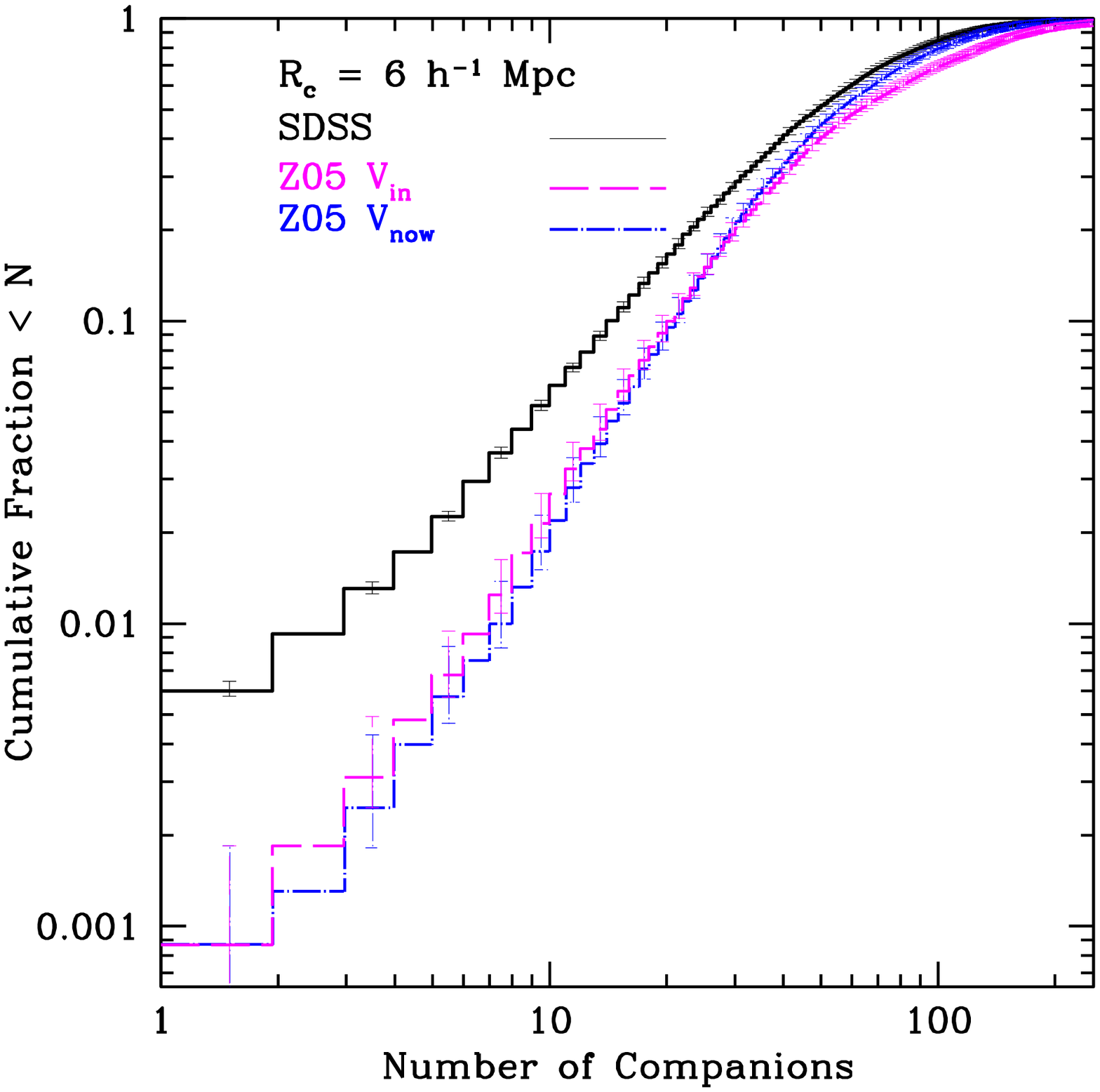} 
\caption{  {\bf  Left:}  Cylinder   counts  for  Z05  $\Vin$  (magenta
long-dashed line),  $\Vnow$ (blue  dot-dashed line), and  SDSS (smooth
solid black  line) within $\Rc$  = 6 $\hMpc$ cylinders.   {\bf Right:}
Cumulative fraction of of galaxies  or halos with fewer than the given
number of companions within $\Rc$ = 6 $\hMpc$ cylinders. }
\label{fig:f9}
\end{figure*}

We can  use the  $\chi^2$ statistic computed  as a summation  over the
companion counts obtained in each  model to assess the ability of each
model to match the SDSS data.  All values quoted are $\chi^{2}$/(dof),
where the  degrees of freedom  (dof) equal the  number of bins  in the
distribution.  For the purposes  of the $\chi^2$ calculation, all bins
have a width  of one companion (note that this  does not correspond to
the binning in Figures~\ref{fig:f7},~\ref{fig:f8}, and ~\ref{fig:f9}).
We  ignore  the  correlation  between different  counts  in  computing
$\chi^2$ and treat this statistic  only as guidance.  We find that the
Z05 $\Vnow$ model produces a $\chi^{2}/$(dof) value of 1.002.  This is
surprising, as  previous studies (such as  \citet{Conroy06}) find that
$\Vin$  tends to  be the  type  of model  that best  matches two-  and
three-point clustering statistics.  In this case, the Z05 $\Vin$ model
has  a $\chi^{2}$/(dof)  value of  131.264, while  the  Millennium and
MPAGalaxies distributions have $\chi^{2}$/(dof) = 1201.758 and 21.721,
respectively.  If we disregard the tail of the distributions, and look
only at  the bins with  more than 0  but fewer than 10  companions, we
find that the Z05 $\Vin$ value drops to 1.819, while $\Vnow$ is 0.297,
implying  that  the Z05  models  are  more  consistent with  the  SDSS
distribution at  the peak.  The Millennium and  MPAGalaxies models are
not   significantly   improved   by   such  excisions   and   have   a
$\chi^{2}$/(dof) value > 1 for all cylinder radii and subsamples.  The
cumulative fraction  of galaxies  with less than  or equal to  a given
number of companions tells a similar story.

For  the  sake  of clarity,  we  do  not  include the  Millennium  and
MPAGalaxies samples  in the remaining figures.  We  do, however, quote
their   $\chi^{2}$/(dof)    values   and   companion    fractions   in
Tables~\ref{tab:t2} and ~\ref{tab:t3}.

The 3  $\hMpc$ cylinder produces  a more problematic  distribution, as
seen  in Figure~\ref{fig:f8}.   While the  general shapes  of  the Z05
distributions look similar to the SDSS result, note that the first bin
is is very  low for both the models.   The $\chi^{2}$/(dof) values for
$\Vin$ and  $\Vnow$ in  this case are  6.008 and  0.683, respectively.
However,   if  we   look  only   at  the   peak  (<   50  companions),
$\chi^{2}$/(dof) for $\Vin$ = 0.494 and $\chi^{2}$/(dof) for $\Vnow$ =
7.379,  favoring the  $\Vin$ model.   On the  right hand  side  of the
figure  the discrepancy  in  the first  bin  is plainly  demonstrated.
While  1.7  $\%$  of the  galaxies  in  the  SDSS  data  have 1  or  0
companions, the simulations have a frequency of less than half that.

It  is interesting  to  note that  the  means of  the SDSS and  simulation
distributions appear  to be consistent  with each other  (the greatest
difference from the  average of the means is  7.12, while the standard
deviation of the means from  the Z05-sized volumes is 7.24), while the
counts-in-cylinders statistics  are not.   The agreement of  the means
may be expected from the  agreement in the correlation function.  This
result  is   another  illustration  of  the   complementarity  of  the
information  contained in  the full  distribution of  companion counts
compared  to either  the correlation  function or  the  mean companion
count.

The  results  for  the  $\Rc$  =  6  $\hMpc$  cylinder  are  shown  in
Figure~\ref{fig:f9}.   The  underprediction   by  the  simulations  of
galaxies  with  few  companions  is  more  pronounced  at  6  $\hMpc$,
extending to galaxies with  up to 20 companions.  The $\chi^{2}$/(dof)
values in  this case for the  whole distribution are  1.047 for $\Vin$
and 0.514 for $\Vnow$. 

Disregarding the first bin brings the Z05 values to $\chi^{2}$/(dof) =
0.944 for $\Vin$ and 0.401 for  $\Vnow$.  Looking at just the peak (20
< Number of  Companions < 110), we get 0.337 and  1.050 for $\Vin$ and
$\Vnow$, respectively. On the other  hand, if we look only at galaxies
with fewer  than 20 companions,  we find $\chi^{2}$/(dof) =  3.292 and
3.800 for $\Vin$  and $\Vnow$.  Again, the Z05  distributions are more
consistent with the SDSS results at the peak than at low densities.

\begin{table*}[t]
\caption{
Results of the $\chi^{2}$ analysis
}
\label{tab:t2}
\begin{center}
\begin{small}
\begin{tabular}{|l|l|l|l|l|}

\hline
$\Rc (\hMpc)$& Z05 $\Vin$& Z05 $\Vnow$& Millennium DM& MPAGalaxies\\
\hline
\hline
\hline

$\Rc$ = 1 & & & & \\
\hline
Full& $\chi^{2}$/(dof) = 131.264& $\chi^{2}$/(dof) = 1.002& $\chi^{2}$/(dof) = 1201.758& $\chi^{2}$/(dof) = 21.721\\
\hline
0 < bin < 10& $\chi^{2}$/(dof) = 1.819& $\chi^{2}$/(dof) = 0.297& $\chi^{2}$/(dof) = 4.629& $\chi^{2}$/(dof) = 4.149\\

\hline
\hline

$\Rc$ = 3 & & & & \\
\hline
Full& $\chi^{2}$/(dof) =  6.008& $\chi^{2}$/(dof) = 0.683& $\chi^{2}$/(dof) = 5.232& $\chi^{2}$/(dof) = 23.441\\
\hline
0 < bin < 50& $\chi^{2}$/(dof) = 0.494& $\chi^{2}$/(dof) = 7.379& $\chi^{2}$/(dof) = 3.011& $\chi^{2}$/(dof) = 2.604\\

\hline
\hline

$\Rc$ = 6 & & & & \\
\hline
Full& $\chi^{2}$/(dof) = 1.198& $\chi^{2}$/(dof) = 0.514& $\chi^{2}$/(dof) = 3.675& $\chi^{2}$/(dof) = 4.349\\
\hline
20 < bin < 110& $\chi^{2}$/(dof) = 0.337& $\chi^{2}$/(dof) = 1.050& $\chi^{2}$/(dof) = 1.769& $\chi^{2}$/(dof) = 1.558\\

\hline

\end{tabular}
\end{small}
\end{center}
\end{table*}



\begin{table*}[t]
\caption{
Fraction of galaxies with fewer than given number of companions
}
\label{tab:t3}
\begin{center}
\begin{scriptsize}
\begin{tabular}{|l|l|l|l|l|l|}

\hline
Number of Companions& SDSS& Z05 $\Vin$& Z05 $\Vnow$& Millennium DM& MPAGalaxies\\
\hline
\hline
\hline

$\Rc$ = 1 $\hMpc$& & & & & \\
\hline
<= 1& 0.352 $\pm$ $^{0.016}$ $_{0.008}$& 0.308 $\pm$ $^{0.006}$ $_{0.004}$& 0.360  $\pm$ $^{0.006}$ $_{0.004}$& 0.391& 0.327\\ 
\hline
<= 3& 0.582  $\pm$ $^{0.018}$ $_{0.010}$& 0.497 $\pm$ $^{0.021}$ $_{0.009}$& 0.595 $\pm$ $^{0.021}$ $_{0.009}$& 0.664& 0.530\\
\hline
<= 5& 0.715 $\pm$ $^{0.019}$ $_{0.013}$& 0.607 $\pm$ $^{0.034}$ $_{0.016}$& 0.726 $\pm$ $^{0.034}$ $_{0.016}$& 0.807& 0.641\\
\hline
<= 10& 0.861 $\pm$ $^{0.021}$ $_{0.016}$& 0.744  $\pm$ $^{0.050}$ $_{0.026}$& 0.870 $\pm$ $^{0.050}$ $_{0.026}$& 0.941& 0.777\\

\hline
\hline

$\Rc$ = 3 $\hMpc$& & & & & \\
\hline
<= 1& 0.017 $\pm$ $^{0.002}$ $_{0.001}$& 0.006 $\pm$ $^{0.003}$ $_{0.001}$& 0.005  $\pm$ $^{0.003}$ $_{0.001}$& 0.005& 0.008\\ 
\hline
<= 3& 0.070  $\pm$ $^{0.006}$ $_{0.005}$& 0.036 $\pm$ $^{0.012}$ $_{0.004}$& 0.035 $\pm$ $^{0.012}$ $_{0.004}$& 0.035& 0.045\\
\hline
<= 10& 0.310 $\pm$ $^{0.020}$ $_{0.012}$& 0.239 $\pm$ $^{0.025}$ $_{0.013}$& 0.258 $\pm$ $^{0.025}$ $_{0.013}$& 0.273& 0.254\\
\hline
<= 50& 0.900 $\pm$ $^{0.024}$ $_{0.018}$& 0.786  $\pm$ $^{0.032}$ $_{0.021}$& 0.887 $\pm$ $^{0.032}$ $_{0.021}$& 0.943& 0.776\\

\hline
\hline

$\Rc$ = 6 $\hMpc$& & & & & \\
\hline
<= 2& 0.002 $\pm$ $^{0.0004}$ $_{0.0003}$& 0.0003 $\pm$ $^{0.0002}$ $_{0.0002}$& 0.0001  $\pm$ $^{0.0002}$ $_{0.0002}$& 0.00007& 0.0003\\ 
\hline
<= 5& 0.009  $\pm$ $^{0.0009}$ $_{0.0008}$& 0.002 $\pm$ $^{0.001}$ $_{0.0006}$& 0.001 $\pm$ $^{0.001}$ $_{0.0005}$& 0.001& 0.004\\
\hline
<= 50& 0.485 $\pm$ $^{0.014}$ $_{0.013}$& 0.378 $\pm$ $^{0.023}$ $_{0.013}$& 0.416 $\pm$ $^{0.023}$ $_{0.013}$& 0.440& 0.369\\
\hline
<= 110& 0.872 $\pm$ $^{0.018}$ $_{0.016}$& 0.719  $\pm$ $^{0.027}$ $_{0.016}$& 0.819 $\pm$ $^{0.027}$ $_{0.017}$& 0.890& 0.706\\
\hline

\end{tabular}
\end{scriptsize}
\end{center}
\end{table*}




\section{Discussion} 
\label{sec:discussion}

The comparison shows  a mismatch between the galaxy  assignment in the
simulations  and   the  SDSS  data,   when  we  look  at   the  entire
distribution.  Here  we test  the robustness of  the result  to simple
changes in the galaxy assignment scheme, and also investigate possible
observational   effects.    \S~\ref{sec:cm},  \S~\ref{sec:vmax},   and
\S~\ref{sec:isogals} detail our results.

\subsection{Color Modeling} 
\label{sec:cm}

One possible reason that the simulations do not match the SDSS data at
large ($\Rc$ = 6 $\hMpc$) scales is  that we do not include any of the
effects of varying  mass-to-light ratios (M/L) in the  models to which
we assign luminosities (the Z05  models and the Millennium dark matter
halos).  It has been demonstrated  that galaxy color is related to the
environment, with galaxies in denser environments having redder colors
and   higher  M/L   \citep{Balogh04,  Hogg04,   Tanaka04,  Weinmann06,
Poggianti06,  Martinez06, Gerke07}.   Here, we  use  the relationships
from \citet{Weinmann06},  who relate the fraction of  early, late, and
intermediate   galaxies   to  host   dark   matter   halo  mass,   and
\citet{Bell01},  who  derive  a  color  dependent  M/L,  where  redder
galaxies are dimmer for a given mass.  To determine the effects of M/L
on our statistic, we use the median color as a function of parent halo
mass from Figure  11 of \citet{Weinmann06} to assign  a $g-r$ color to
each subhalo. These colors were then used with the color-dependent M/L
for  SDSS  bandpasses from  \citet{Bell03}  to  calculate an  adjusted
luminosity or  an effective  stellar mass.  We  then found  a weighted
number  density from these  adjusted luminosities,  which was  used to
determine the actual luminosity from the SDSS luminosity function.

The color corrected Z05 $\Vin$  results for the 6 $\hMpc$ cylinder are
shown  in  the   left  panel  of  Figure~\ref{fig:f10}.   This
analysis  uses the largest  possible contrast  between late  and early
types, therefore  causing the largest  shift.  While the shift  in the
distribution  is in the  correct direction  (more galaxies  with fewer
companions),  it only  accounts for  a  small part  of the  difference
between  the simulation and  observational data.   The lack  of simple
color modeling is not the  cause of this discrepancy.  The mismatch of
the MPAGalaxies  sample, which  does include color  modeling, further
supports this result.

\subsection{Varying the $\Vmax$ Cut} 
\label{sec:vmax}

\begin{figure*}[t!]
\centering
\epsscale{1.0} 
\plottwo{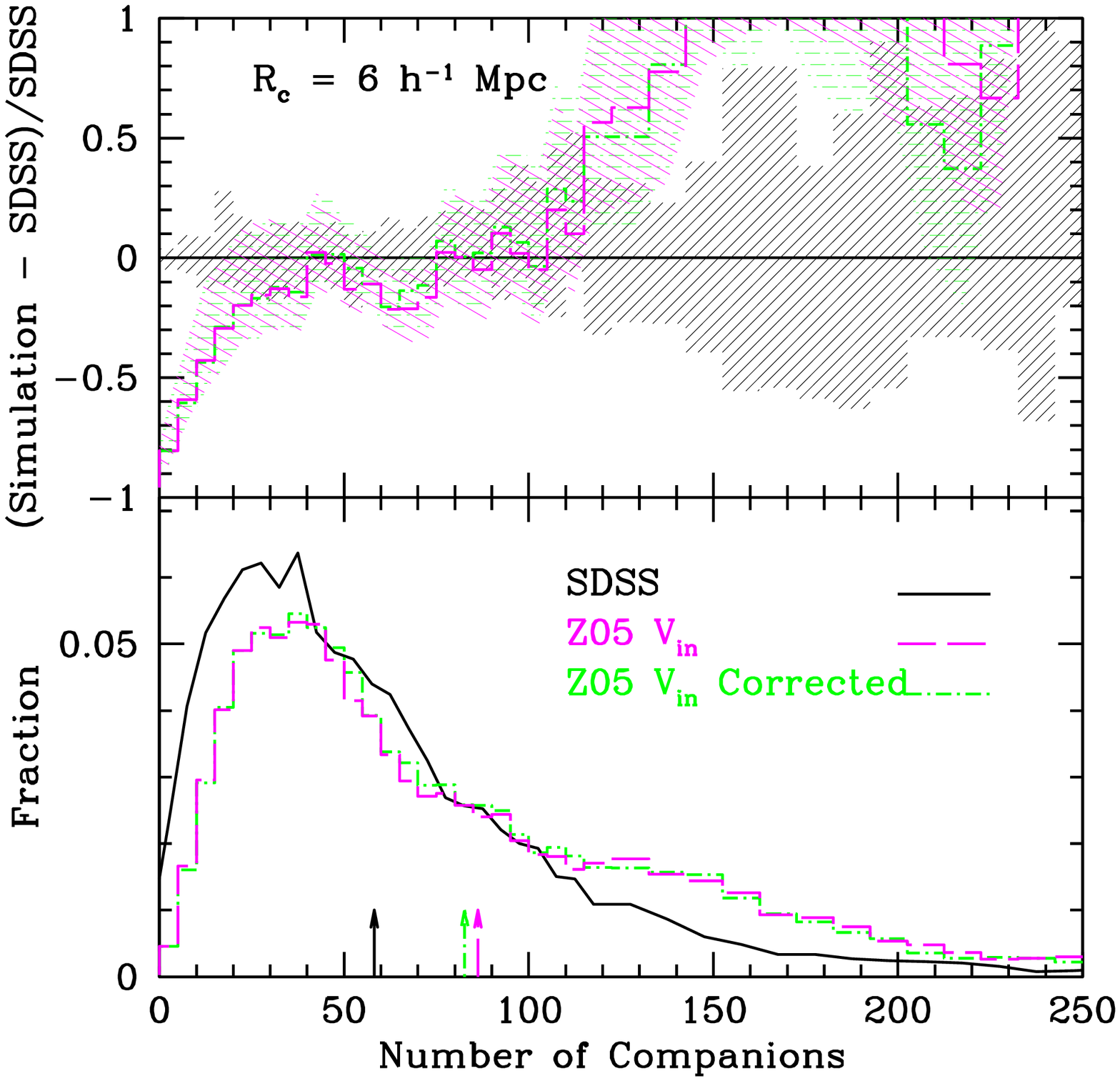} {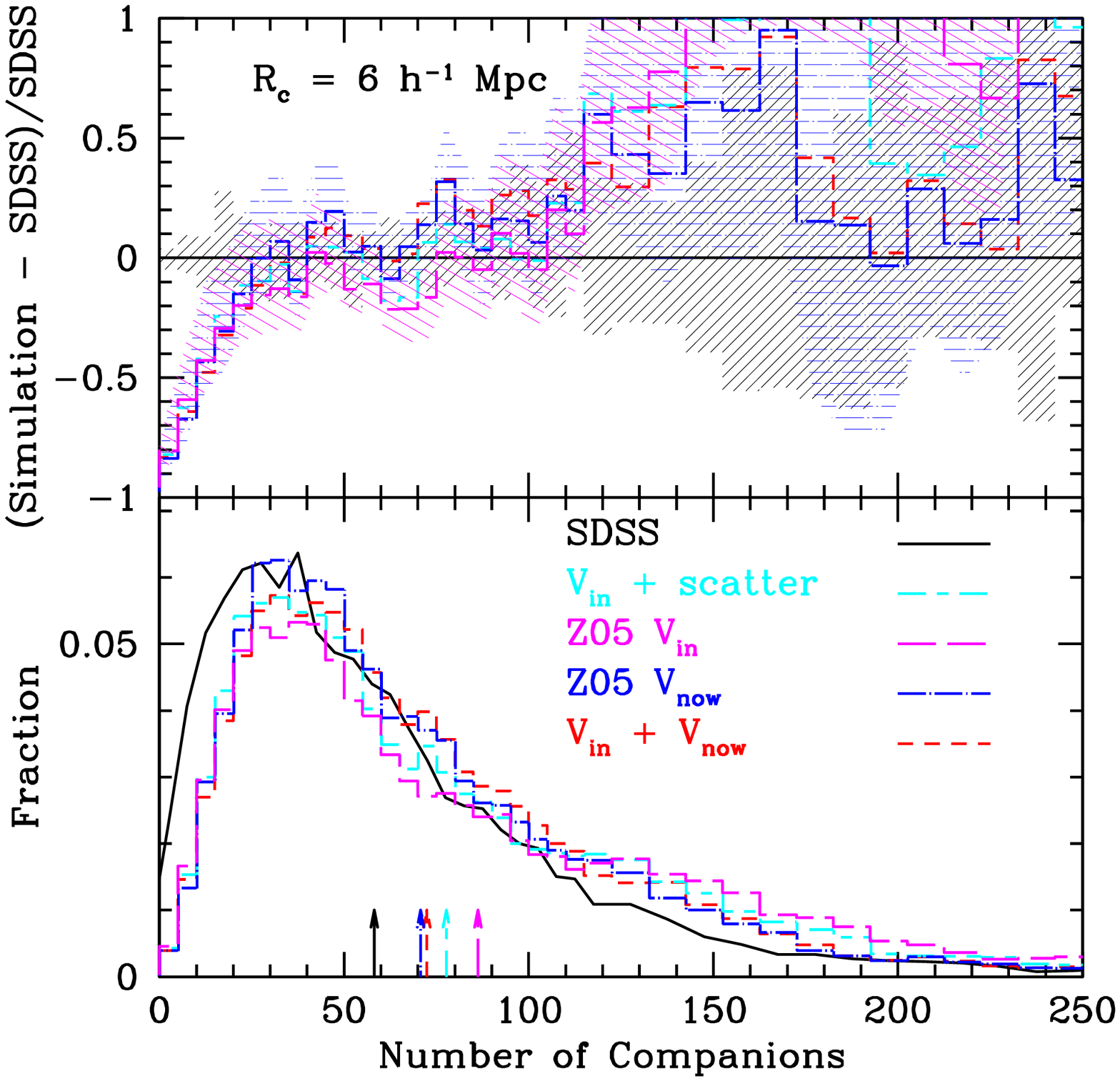}
\caption{  {\bf Left:} Cylinder  counts for  SDSS (smooth  solid black
line),  Z05 $\Vin$  (magenta long-dashed  line), and  Z05  $\Vin$ with
basic  color correction  (green  short-dashed line)  within 6  $\hMpc$
cylinders. {\bf  Right:} Cylinder counts for SDSS  (smooth solid black
line),  Z05  $\Vin$  (magenta  long-dashed line),  Z05  $\Vnow$  (blue
dot-dashed   line),  a   combination  of   $\Vin$  and   $\Vnow$  (red
short-dashed line), and $\Vin$ with  scatter in the $\Vmax$ cut ((cyan
long-dashed--short-dashed line) within $\Rc$ = 6 $\hMpc$ cylinders.}
\label{fig:f10}
\end{figure*}

To  further explore  possible  explanations for  the  mismatch in  the
distributions at  large scales, we try  two other models  with the Z05
data.  First, we  vary the $\Vin$ model halos  being detected near the
$\Vmax$ cutoff  by decreasing  the $\Vmax$ cutoff  by a  random number
between  0 and  25  km s$^{-1}$  for  host halos,  and increasing  the
$\Vmax$ cutoff  by a random number  between 0 and 100  km s$^{-1}$ for
subhalos before  calculating the number  density. This is a  step away
from the monotonic luminosity assignment that we have been using, and,
by  allowing more  small host  halos, might  increase the  fraction of
isolated  galaxies.  The  resulting distribution,  shown in  the right
panel of  Figure~\ref{fig:f10}, falls  in between the  original $\Vin$
and $\Vnow$  distributions.  While it  certainly changes the  shape of
the  distribution,  the $\chi^{2}$/(dof)  value  (3.292  for the  full
distribution) does not significantly  improve on the original results,
or increase the fraction of halos with < 20 companions.

The right  panel of  Figure~\ref{fig:f10} also shows  the distribution
for a second model.  Here we use the $\Vin$ velocities for subhalos in
hosts with  $\Vmax$ <  $10^{13}$ $\Msun$ and  $\Vnow$ for  subhalos in
hosts with $\Vmax$ >=  $10^{13}$ $\Msun$.  This procedure assumes that
galaxies in  large halos  are more likely  to be stripped  of luminous
matter than galaxies  in small halos.  This distribution  results in a
$\chi^{2}$/(dof) = 2.800.  Once again, this does not  correct the lack
of isolated halos in the simulations.

\subsection{Isolated Galaxies} 
\label{sec:isogals}

The consistent feature in the cylinder counts distribution for all the
of the models is the relative paucity of isolated galaxies compared to
the observed frequency  of isolated galaxies.  To verify  that this is
not an artifact  of the survey footprint, we  examine a possible cause
in our  analysis of the SDSS data  for the $\Rc$ =  6 $\hMpc$ cylinder
counts,  where the  effect is  most pronounced.   First, we  check the
images  of  the twenty-one  galaxies  in  SDSS  flagged as  having  no
companions.  We find that ten are  within 6 $\hMpc$ of an edge.  While
we do correct for edge effects  in our analysis, we also recompute the
fraction  of isolated  galaxies without  those ten  as  a conservative
estimate.   Adopting  the  conservative  estimate of  eleven  isolated
galaxies in the SDSS distribution, we expect to find eighteen isolated
halos in  the $\Vin$  sample.  We only  find seven.   Assuming Poisson
statistics,  the  probability that  we  should  find  so few  isolated
galaxies  in the  $\Vin$ sample  is $\approx  2 \times  10^{-3}$ (this
compares to  $\sim 10^{-9}$  if all twenty  one isolated  galaxies are
included).  The probabilities for the other models are all lower by at
least  one order  of magnitude.   We conclude  that the  difference in
isolated  galaxy  fraction  is  not  caused by  unaccounted  for  edge
effects.



\section{Conclusions} 
\label{sec:conc}

We  have explored the  counts-in-cylinders statistic,  and used  it to
compare galaxy environments in  the SDSS with environments measured in
several  models  for the  galaxy  distribution  based  on dark  matter
simulations.    We  show  that   this  statistic   provides  different
information than the two-point function  alone; it is possible for two
catalogs  to   have  similar  two-point   correlation  functions,  but
companion distributions with very different shapes.

Our primary results are as follows.

\begin{enumerate}
\item There  is a  large scatter  in the number  of companions  a dark
  matter halo of a given host mass or halo occupation has within a set
  cylinder.  The  counts-in-cylinders statistic  is limited as  a tool
  for determining the host halo mass of a galaxy.

\item We considered several models  for assigning galaxies to the dark
  matter distribution, including models based on abundance matching to
  dark  matter substructures as  well a  semi-analytic model  from the
  Millennium   simulation.   Each   of   these  models   significantly
  underpredicts  the number of  galaxies with  very few  companions on
  $\Rc$ = 3 and 6 $\hMpc$ scales.

\item  While  none  of  the  simulations or  models  examined  have  a
  counts-in-cylinders  distribution that  is consistent  with  that of
  SDSS  data,  the two  abundance  matching  models  (Z05 $\Vnow$  and
  $\Vin$)  have  similar  distributions   when  the  first  few,  very
  discrepant, bins  (corresponding to the most  isolated galaxies) are
  ignored.


\item The counts-in-cylinders test fails for models that match the
two-  and three-point correlation functions, highlighting its utility as a 
diagnostic.

\end{enumerate} 
We have tested the robustness of these results to a series of possible
systematic errors.  Simple  changes to the color assignment  or to the
scatter model  in the  abundance matching approach  do not  change the
conclusions.  We have accounted  for known effects in the completeness
of the  SDSS NYU-VAGC data.  In addition,  is its hard to  see how any
small  scale  incompleteness would  explain  the  discrepancy seen  on
several Mpc scales.

Our results indicate that some  observed galaxies in the real universe
are significantly more isolated than  any halos of comparable size. It
does  not  appear that  the  discrepancy  we  have identified  in  the
counts-in-cylinders  can  be  easily  resolved with  a  standard  halo
occupation approach that assumes that all of a galaxy's properties are
set by the  mass of its host halo.  In any  case, this mismatch merits
further study.


\acknowledgments

The ART simulation  was run  on the Seaborg  machine at  Lawrence Berkeley
National  Laboratory (Project  PI:  Joel Primack).   We thank  Anatoly
Klypin for running the simulation  and making it available to us.  The
Millennium  Simulation  databases  used  in  this paper  and  the  web
application providing  online access to them were  constructed as part
of  the activities  of the  German Astrophysical  Virtual Observatory.
Funding for the  Sloan Digital Sky Survey (SDSS)  has been provided by
the Alfred  P. Sloan  Foundation, the Participating  Institutions, the
National  Aeronautics and Space  Administration, the  National Science
Foundation,   the   U.S.    Department   of   Energy,   the   Japanese
Monbukagakusho,  and the  Max Planck  Society.  The  SDSS Web  site is
http://www.sdss.org/.   The  SDSS  is  managed  by  the  Astrophysical
Research  Consortium (ARC)  for the  Participating  Institutions.  The
Participating  Institutions are The  University of  Chicago, Fermilab,
the Institute  for Advanced Study, the Japan  Participation Group, The
Johns  Hopkins   University,  Los  Alamos   National  Laboratory,  the
Max-Planck-Institute  for Astronomy  (MPIA),  the Max-Planck-Institute
for  Astrophysics (MPA),  New Mexico  State University,  University of
Pittsburgh, Princeton University, the United States Naval Observatory,
and the University of Washington.

HDB, EJB, JSB are supported by  the Center for Cosmology at UC Irvine.
JSB  is supported  by the  National Science  Foundation  (NSF) through
grant AST-0507916. JCB was supported by the Center for Cosmology at UC
Irvine, and  the NSF through grant  AST-0507916 for a  portion of this
work, and is currently supported by the University of Arkansas. ARZ is
supported  by the  University of  Pittsburgh, the  NSF  through Grants
AST-0602122 and AST-0806367, and by  the US Department of Energy.  RHW
was supported in part by  the U.S. Department of Energy under contract
number  DE-AC02-76SF00515 and  by  a Terman  Fellowship from  Stanford
University.

\bibliography{ms}

%
%
\newpage

%
%
%
%
%

%
%
%
%
%

%
%

%
%

%
%
%

%
%
%
%
%
%

%
%
%
%
%
%

%
%
%
%
%
%

%
%
%

%
%
%

%
%
%
%
%
%
%

%
%
%

%
%
%

\end{document}